\documentclass[aps,prl,twocolumn,superscriptaddress,floatfix,nofootinbib,showpacs,longbibliography]{revtex4-2}
\usepackage[left=15mm,right=15mm,top=20mm,bottom=35mm,paper=a4paper]{geometry}
\usepackage[colorlinks=true, linkcolor=blue, citecolor=blue, urlcolor=blue]{hyperref}
\usepackage{amsthm} 

\newtheorem{theorem}{Theorem}

\newtheorem{lemma}{Lemma}

\usepackage{graphicx}  
\usepackage{subfigure}
\usepackage{multirow}
\usepackage{footnote}
\usepackage{ragged2e} 

\usepackage{parskip}
\usepackage[T1]{fontenc}
\usepackage{dcolumn}   

\usepackage{bm}        
\usepackage{amsfonts}  
\usepackage{amsmath}   
\usepackage{amssymb}   

\usepackage{hyperref}   

\usepackage{xcolor}
\usepackage{caption}

\usepackage{subcaption}
\usepackage{physics}

\usepackage{natbib} 
\usepackage{amsmath,amssymb,amsthm,bm,amsfonts,mathrsfs,bbm,dsfont} 

\newcommand{\Id}{\mathds{1}}

\begin{document}

\title{Precision bounds for bosonic quantum batteries}

\author{Beatriz Polo-Rodríguez}
\thanks{beatriz.polo@icfo.eu}
\affiliation{ICFO-Institut de Ciencies Fotoniques, The Barcelona Institute of Science and Technology, Av. Carl Friedrich Gauss 3, 08860 Castelldefels (Barcelona), Spain.}

\author{Federico Centrone}
\thanks{fe.centrone@gmail.com.}
\affiliation{ICFO-Institut de Ciencies Fotoniques, The Barcelona Institute of Science and Technology, Av. Carl Friedrich Gauss 3, 08860 Castelldefels (Barcelona), Spain.}
\affiliation{Universidad de Buenos Aires, Instituto de Física de Buenos Aires (IFIBA), CONICET,
Ciudad Universitaria, 1428 Buenos Aires, Argentina.}

\begin{abstract}  

We study precision charging in bosonic quantum batteries under a finite-energy constraint, using the \emph{signal-to-noise ratio} (SNR) of delivered excitations as an operational metric directly tied to the energy measured at a load. At the state level, we derive a \emph{classical} bound whose violation is equivalent to antibunching and certifies non-classicality, and a \emph{Gaussian} bound whose violation certifies non-Gaussianity under fixed temperature and energy-input constraints. We identify experimentally accessible non-Gaussian families that surpass this Gaussian bound at finite temperature, thereby establishing non-Gaussianity as a resource for enhanced charging precision. Finally, we introduce a linear photodetection model which, under standard linear-response assumptions, propagates these bounds to the photocurrent level and enables both witnesses to be evaluated solely from electrical statistics. Together, these results provide a realistic route to demonstrating an operational quantum advantage—defined as surpassing classical and Gaussian precision bounds—in a thermodynamically motivated energy-conversion task, with plausible near-term applications to the precision charging of fragile nanoscopic loads.

\end{abstract}

\maketitle 

\section{I. Introduction}
While originally motivated by the industrial need to optimize engine performance, thermodynamics has produced some of science's deepest and most interconnected concepts, ranging from cosmology to information theory to biology \cite{goldstein1995refrigerator}. Quantum thermodynamics \cite{binder2018thermodynamics} has the potential to share the same success as its classical counterparts, in a moment where energetic optimization is a fundamental demand and where quantum technologies are receiving notable attention \cite{auffeves2022quantum}. Quantum batteries are one of the most trending topics in the community \cite{campaioli2024colloquium,Konar_2024,QUACH20232195,Camposeo_2025,shi2025quantumchargingadvantagemultipartite}, demonstrating the promise of an energetic quantum advantage in terms of charging power. While being a concept of fundamental interest, power is not a significant figure of merit for practical scenarios, because of the low energetic scales of quantum systems and the resources required to prepare them \cite{kurman2025quantumcomputationquantumbatteries}. In addition, while quantum systems can store work in many ways, studies of the practical extraction and use of such energy are limited \cite{myers2022quantum}.\\
While much of the recent quantum-battery literature emphasizes large energies or charging powers, the edge of quantum mechanics has historically been associated with precision, as demonstrated by quantum metrological experiments \cite{Liu2021,yin2023experimental,jia2024squeezing}. Precision is also a prominent topic in thermodynamics since its statistical formulation. In a future fueled by quantum technologies, it is hard to imagine an array of cold atoms powering a laptop or anything similar. However, to power emerging nanotechnologies, it will be necessary to account for quantum fluctuations \cite{lahaye2004approaching}. Fragile microscopic devices require sufficiently precise energy delivery to produce useful work, while classical fluctuations risk either damaging the device or failing to charge it \cite{ussia2024unlocking}. Examples of externally powered micro/nano-actuators include systems driven by light pulses, magnetic fields, or acoustic waves \cite{mishra2012review,giri2021brief,zhou2021magnetically}. The natural language to describe these processes in the quantum regime is the continuous-variables (CV) formalism \cite{serafini2023quantum}. In what follows, we adopt the quantum-optics nomenclature, but our framework directly applies to any bosonic field (e.g., phonons).\\
In this work, we analyze \emph{precision charging} for bosonic quantum batteries under a finite-energy constraint, using the \emph{signal-to-noise ratio} (SNR) of delivered excitations as an operational metric tied to what a harvester actually measures. On the theory side, we establish two bounds that promote this metric to \emph{witnesses}: (i) a \emph{classical} bound (Theorem~\ref{thm:A}) that is equivalent to antibunching and certifies non-classicality, and (ii) a \emph{Gaussian bound} (Theorem~\ref{thm:B}) at fixed temperature and energy input, whose violation certifies non-Gaussianity. We then identify explicit, experimentally motivated non-Gaussian families that surpass the Gaussian bound in relevant parameter regimes. This complements fluctuation-minimizing constructions in charging \cite{Friis_2018} and SNR-based analyses in specific models \cite{rinaldi2024reliablequantumadvantagequantum} by giving a finite-energy Gaussian bound directly in the conversion metric and by identifying concrete non-Gaussian resources that can, in principle, be certified from electrical readout statistics under standard linear photodetection assumptions.\\
From a quantum-information perspective, our bounds endow the resource of non-Gaussianity in continuous-variable systems with a task-oriented witness: any state whose conversion SNR exceeds the Gaussian bound $\Gamma_G(\nu,\epsilon)$ provides strictly stronger performance than all Gaussian states under the same constraints. This connects non-Gaussianity—identified as a key resource for continuous-variable quantum information processing—to a thermodynamically motivated energy-conversion task, in close analogy to resource-theoretic approaches to non-Gaussianity and other nonclassical resources in CV systems.

In the remainder of this work, Sec.~\hyperref[secII]{II} introduces the bosonic battery–harvester model and defines the conversion SNR $\Gamma$ under a finite-energy constraint. Sec.~\hyperref[secIII]{III} derives the classical and Gaussian precision bounds (Theorems~A and B), establishes the finite-energy Gaussian bound $\Gamma_G(\nu,\epsilon)$, and analyzes non-Gaussian families and multimode interferometric enhancements. Sec.~\hyperref[secIV]{IV} discusses photodiode-based implementations and realistic operating regimes, while Sec.~\hyperref[sec:detectable_precision]{V} propagates the bounds through a lossy/noisy detection model and benchmarks the detectable precision against thermodynamic-uncertainty limits. We conclude in Sec.~\hyperref[sec:discussion]{VI} with a discussion and outlook.

\section{II. Bosonic batteries} \label{secII}
\begin{figure}
    \centering
\includegraphics[width=0.5\textwidth]{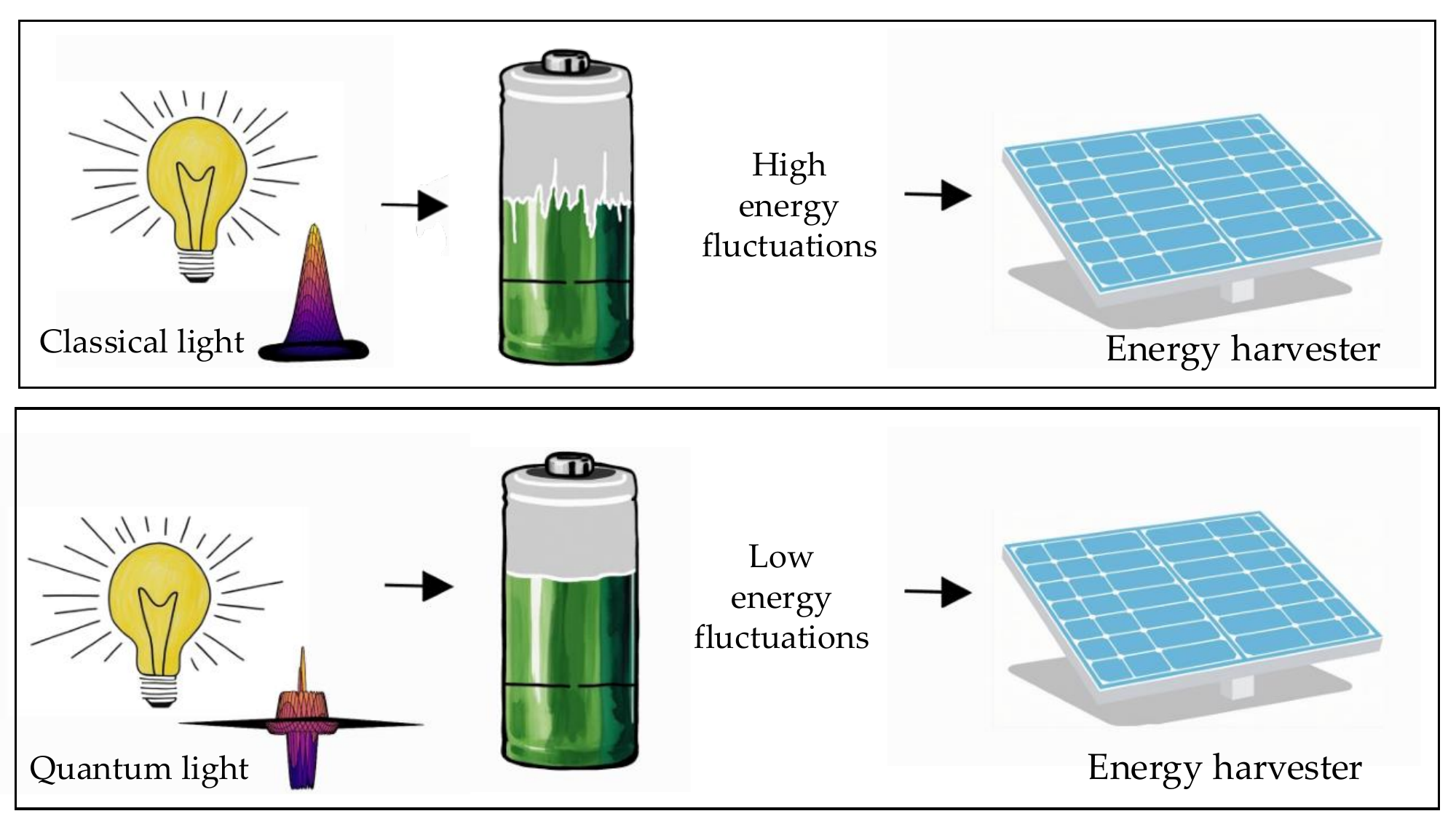}
    \caption{\raggedright\justifying Scheme of the charging precision in a photonic-based battery. Energy fluctuations are attenuated when employing non-classical light.}
\label{fig:batteryScheme}
\end{figure}

A quantum battery is defined as a system whose internal
Hamiltonian $H_0$ has non-degenerate energy levels\footnote{In fact, a more relaxed condition of a partially degenerated spectrum, i.e. such that the eigenvalues satisfy the relation $\epsilon_k \leq \epsilon_{k+1}$ as long as the bandwidth $\epsilon_{\text{max}} - \epsilon_\text{min}$ is non-zero, is sufficient for the definition of a quantum battery \cite{campaioli2024colloquium}.}, such that energy can be temporarily stored by performing unitary operations $U$ on an initial thermal state  $\rho_0$ to prepare the system in some excited state $\rho = U \rho_0 U^\dagger $. The energy stored in the battery during this process will afterwards act as the external source for some other device (see Fig.\ref{fig:batteryScheme}), referred to in this work as \textit{energy harvester}, capable of transforming the battery's potential into useful work (e.g. to power some nanotechnological task \cite{jaliel2019experimental,culhane2022extractable,meng2024quantum}). In the following, we will assume the battery to be a free propagating electromagnetic field (i.e., a light pulse), however, the same analysis works for any oscillatory system, such as harmonic excitations in crystals. \\

As for the energy harvester, examples include photo-diodes, piezoelectric crystals, or quantum dots \cite{phillips2021energy}. We model the harvester as a device producing an electrical response proportional to the excitation number $N=\sum_i a_i^\dagger a_i$ of the external field (with bosonic ladder operators and a discrete mode index for clarity). For pulsed operation at repetition rate $f$, the mean electrical output is proportional to $\langle N\rangle$ (a detailed device model is given later), so optimizing the \emph{precision} of the delivered energy reduces to controlling the mean and fluctuations of $N$. \\

Initially ($t<0$), the optical field is in a thermal passive state at temperature $T$ for the free Hamiltonian $H_0=\sum_i \omega_i(n_i+\tfrac{1}{2})$, i.e.
\begin{equation}
 \rho_0=\frac{e^{-H_0/k_B T}}{\Tr[e^{-H_0/k_B T}]},\qquad 
 N_0=\Tr[N\,\rho_0].
\end{equation}
For $t>0$ the state evolves under $H(t)=H_0+V(t)$, where a charging potential $V(t)$ modifies the photon number to charge the battery. We write the protocol unitary as $U=\mathcal{T}\exp\!\big[-\tfrac{i}{\hbar}\int_0^\tau V(t)\,dt\big]$ and set $V(0)=V(\tau)=0$ for a finite charging window. In the theoretical analysis we work with the number operator and neglect spectral structure; experimental tests fold the detected spectrum into the effective responsivity (see “Implementability”).

In practice $V(t)$ is implemented by a small set of experimentally standard blocks:
\begin{align}
\label{eq:VG}
V_G(t)&=\hbar\!\left[\xi(t)\,a^\dagger+\xi^*(t)\,a\right]+\frac{\hbar}{2}\,r(t)\!\left(e^{i\phi}a^2+e^{-i\phi}a^{\dagger 2}\right),\\[3pt]
\label{eq:VNG}
V_{\mathrm{NG}}(t)&=\hbar K(t)\,(a^\dagger a)^2\ +\ \hbar\chi^{(2)}(t)\!\left(a b + a^\dagger b^\dagger\right),
\end{align}
where $V_G$ generates displacement and squeezing (Gaussian unitaries) and $V_{\mathrm{NG}}$ covers Kerr dynamics and heralded photon addition via an ancilla mode $b$ (conditioning on $b$’s detection implements $(a^\dagger)^m$). In the numerical study, we take the pulses as piecewise-constant, which reproduces the “instantaneous on/off” idealization used in the figures. These blocks suffice to prepare the Gaussian families (coherent/squeezed thermal) and the non-Gaussian families (photon-added, cat/kitten) analyzed below.

The mean increase in excitations and its fluctuations are
\begin{equation}
 \delta N=\Tr\!\big[N(\rho-\rho_0)\big],\qquad
 \Delta N=\sqrt{\Tr[N^2\rho]-\Tr[N\rho]^2},
\end{equation}
with $\rho=U\rho_0U^\dagger$. Our figure of merit is the \textit{signal-to-noise} ratio (SNR)
\begin{equation}
 \Gamma=\frac{\delta N}{\Delta N},
\end{equation}
evaluated under an energy bound $\delta N\le \epsilon$.
In the next section, we analyze how classicality (and Gaussianity) restrict the attainable values of this ratio, thereby allowing us to identify the presence of genuine quantum resources whenever these limits are surpassed.

\section{III. Results}  \label{secIII}
\subsection{A. Classical bound}
We say a state is \emph{classical} if it admits a non-negative Glauber--Sudarshan $P$ representation; coherent states and their mixtures are classical, whereas sub-Poissonian light (Mandel $Q<0$) is non-classical. In our context, classical mixtures of coherent states have Poissonian or super-Poissonian number statistics, so reducing photon-number fluctuations below the Poisson level (at fixed mean) is a signature of non-classicality. The standard second-order correlation $g^{(2)}(0)$ \cite{PhysRev.130.2529} relates to $\Gamma$ via
\begin{equation}\label{g_intermsof_gamma}
 g^{(2)}(0)=\left(\Gamma+\frac{N_0}{\Delta N}\right)^{-2}-\frac{1}{\langle N\rangle}+1,
\end{equation}
and $g^{(2)}(0)<1$ is the necessary and sufficient criterion for antibunching \cite{PhysRevA.90.063824}.

\medskip
\begin{theorem}\label{thm:A}
For any \emph{classical} output state, the following bound holds:
\begin{equation}\label{eq:ClassicalWitness}
 \left(\Gamma+\frac{N_0}{\Delta N}\right)^{\!2}\ \le\ \langle N\rangle,
\end{equation}
equivalently $g^{(2)}(0)\ge 1$ via Eq.~\eqref{g_intermsof_gamma}. Hence any observed violation of \eqref{eq:ClassicalWitness} (equivalently, $g^{(2)}(0)<1$) certifies \emph{non-classical} (sub-Poissonian) light in our precision metric.
\end{theorem}
\medskip
Depending on whether the mean photon number is smaller, equal, or greater than its variance, the statistics are termed sub-Poissonian, Poissonian, or super-Poissonian, respectively. Poissonian and super-Poissonian light admit a semi-classical description, whereas sub-Poissonian sources are non-classical and desirable for precision tasks because they suppress photon-number fluctuations \cite{Ann2019, Berchera_2019}.

\subsection{B. Gaussian limitations}
States that can be engineered through Gaussian unitary operations have many desirable properties. They are easy to implement with a high degree of experimental control in state-of-the-art quantum photonics hardware \cite{brask2022gaussianstatesoperations,Zhang2022}. From a theoretical perspective, the properties and dynamics of Gaussian states can be efficiently simulated with a classical computer, and—despite their infinite dimensionality—admit a compact symplectic description. Nevertheless, Gaussian unitaries are limited. In the context of quantum batteries, we show they are suboptimal resources for precision charging \cite{rinaldi2024reliablequantumadvantagequantum}.

For simplicity, we first restrict to the single-mode scenario with an energy-input constraint $\delta N\le \epsilon$; the multimode generalization is discussed in \hyperref[subsec:multimode]{Section~III.C}.

The harvested-energy SNR of a single-mode Gaussian state $\rho$ is
\begin{equation}\label{gamma}
    \Gamma (\rho)= \frac{N(\rho) - N(\rho_0)}{\Delta N (\rho)}
 = \frac{ \frac{1}{4} (\operatorname{Tr}(\sigma) - 2) +  \lVert \bm{\alpha}\rVert ^2 - N(\rho_0)}{\sqrt{\bm{\alpha}^T \sigma \bm{\alpha}  + \frac{1}{8}[\operatorname{Tr}(\sigma^2) -2 ]}}\,,
\end{equation}
where $\bm{\alpha}=\frac{1}{\sqrt{2}}(\langle x\rangle,\langle p\rangle)^T$ is the displacement, $\sigma$ the covariance matrix, and $\rho_0$ the associated passive thermal state with
\begin{equation}\label{nth}
    N_0 = \tfrac{1}{2}(\nu-1), \qquad \nu=\coth\!\big(\tfrac{\omega}{k_BT}\big).
\end{equation}
Writing the squeezing by a real parameter $0<z\leq 1$ (so that $z$ and $1/z$ are the quadrature variances in shot-noise units), the numerator and denominator of \eqref{gamma} read
\begin{equation}
    N (\rho) -  N (\rho_0) = \frac{1}{4} \nu \!\left(z + \frac{1}{z} -2 \right) + \lVert \bm{\alpha} \rVert ^2,
\end{equation}
\begin{equation} 
    \Delta N = \sqrt{ \frac{1}{8} \nu^2 \!\left(z^2 + \frac{1}{z^2} \right) -\frac{1}{4}  +  \nu \!\left(\alpha_1^2 z+\alpha_2^2 \frac{1}{z}\right)}\,.
\end{equation}

Thus, for fixed thermal fluctuations $\nu$, $\Gamma$ is fully determined by $(z,\bm{\alpha})$ (no phase dependence beyond aligning $\bm{\alpha}$ to the squeezed/anti-squeezed axes).

For coherent states ($z=1$),
\begin{equation}\label{snr_coherent}
    \Gamma (\rho_{\text{coh}}) = \frac{\lVert \bm{\alpha} \rVert ^2}{\sqrt{ \frac{1}{4} (\nu^2-1)  +  \nu \lVert \bm{\alpha} \rVert ^2 }},
\end{equation}
and the constraint $\lVert\bm{\alpha}\rVert^2=\delta N\le \epsilon$ implies
\begin{equation}\label{bound_coherent}
    \Gamma (\rho_{\text{coh}}) \le \sqrt{\epsilon}.
\end{equation}
If only squeezing is applied to a thermal seed ($\bm{\alpha}=0$), one finds
\begin{equation}\label{bound_squeezed}
    \Gamma(\rho_{\text{sq}}) \le 1,
\end{equation}
since
\[
\underbrace{\tfrac{1}{4}\nu \!\left(z + \tfrac{1}{z}-2\right)}_{N(\rho)-N(\rho_0)}\le
\underbrace{\sqrt{\tfrac{1}{8}\nu^2 \!\left(z^2+ \tfrac{1}{z^2}\right) -\tfrac{1}{4}}}_{\Delta N}
\qquad \forall z\in(0,\infty).
\]

\medskip
\begin{theorem}\label{thm:B}
Consider the fixed thermal parameter $\nu$ (temperature $T$) and the energy-input constraint $\delta N\le \epsilon$. Defining the \emph{Gaussian bound} as
\begin{equation}\label{eq:GaussianWitness}
 \Gamma_G(\nu,\epsilon)\;:=\;\max_{\text{Gaussian }\rho}\ \Gamma(\rho)\quad\text{subject to } \delta N(\rho)\le \epsilon,
\end{equation}
where $\Gamma(\rho)$ is given by Eq.~\eqref{gamma}, and all Gaussian optimizations are over displacement, single-mode squeezing and passive interferometers; the latter do not change $\Gamma$ (see \hyperref[app:multimode]{Appendix~G}). Then every Gaussian state obeys
\[
 \Gamma(\rho_G)\ \le\ \Gamma_G(\nu,\epsilon).
\]

Consequently, any observation of $\ \Gamma\!>\!\Gamma_G(\nu,\epsilon)\ $ certifies that the state is \emph{non-Gaussian} (under the same $(\nu,\epsilon)$ constraints).
\end{theorem}

\begin{figure}[h!]
    \centering
    \includegraphics[width=\linewidth]{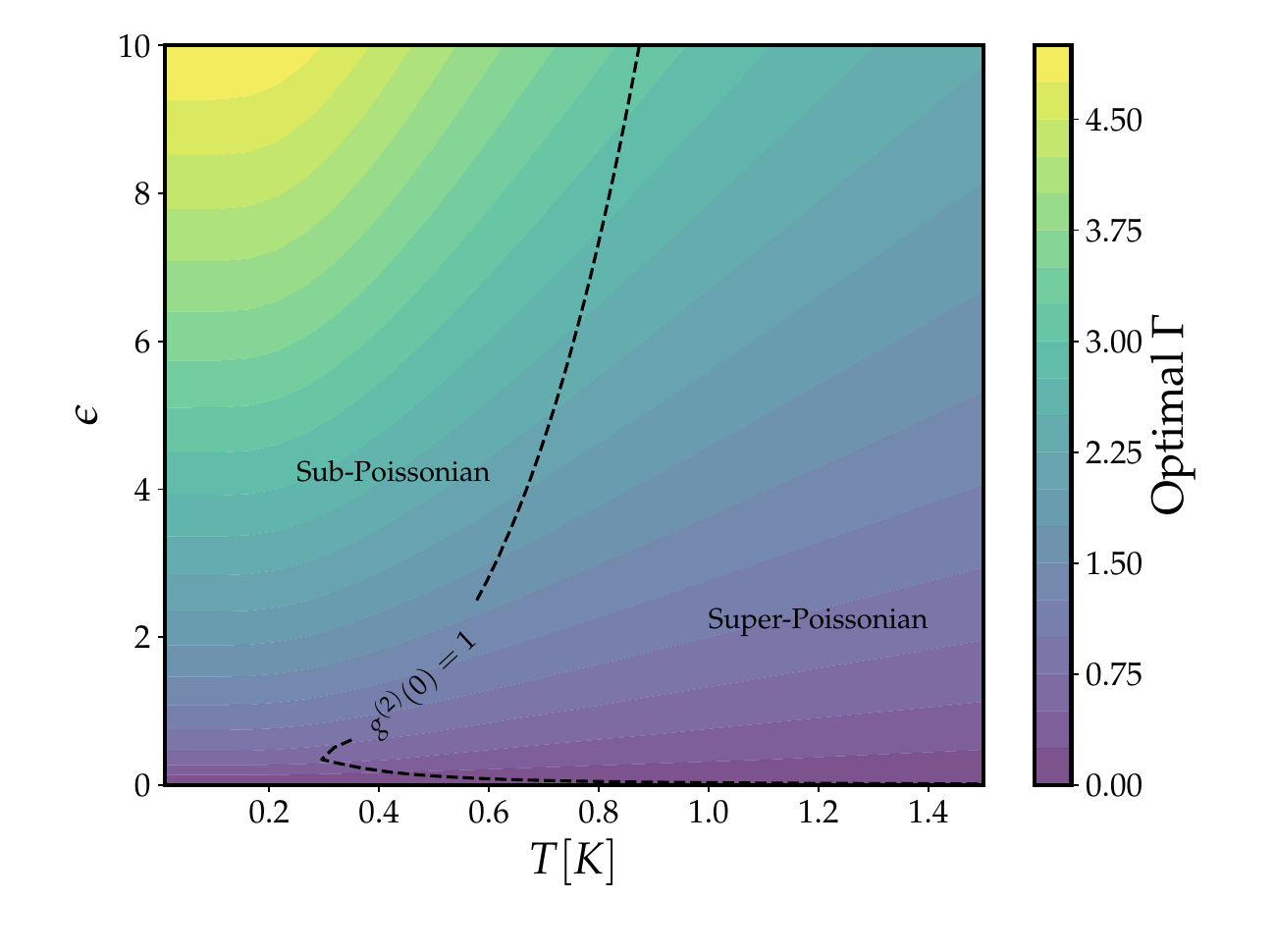}
    \caption{ \raggedright\justifying Maximum photon-number SNR attainable for \emph{Gaussian} states, $\Gamma_G(\nu,\epsilon)$, as a function of temperature (via $\nu$) and maximum energy input $\epsilon$. The dashed line marks $g^{(2)}(0)=1$ (classical/non-classical boundary).}
    \label{lagrange_opt_snr}
\end{figure}

\medskip
Squeezed-only batteries perform worse than coherent states once the state contains more than one photon. The optimal Gaussian SNR at a given temperature and fixed ergotropic constraint $\epsilon$ is therefore obtained by \emph{combining} displacement and squeezing to minimize $\Delta N$ at fixed $\delta N$. We solve this constrained optimization via Lagrange multipliers (see details in \hyperref[app: optimal_gaussian]{Appendix~B}). Figure~\ref{lagrange_opt_snr} shows $\Gamma_G(\nu,\epsilon)$ across temperatures and energy budgets; in the highly squeezed regime an accurate analytical approximation is
\begin{equation}
    \Gamma_{G}(\nu,\epsilon)\ \approx\ \frac{8\epsilon}{\nu^2\!\left[\,3\Big(\frac{4\epsilon}{\nu}+2\Big)^{\!\frac{2}{3}}-2+\Big(\frac{4\epsilon}{\nu}+2\Big)^{\!-\frac{2}{3}} \right] }.
\end{equation}

We observe that by combining squeezing and displacement, one can surpass the \emph{individual} coherent/squeezed bounds \eqref{bound_coherent}–\eqref{bound_squeezed}, and yield optimal Gaussian states lying in the sub-Poissonian region ($g^{(2)}(0))<1$. In the next section we explore how genuinely \emph{non-Gaussian} states can exceed $\Gamma_G(\nu,\epsilon)$, thereby realizing a precision advantage beyond the entire Gaussian sector.

\subsection{C. Non-Gaussian states}
\begin{figure}[h!]
    \centering
    \includegraphics[width=\linewidth]{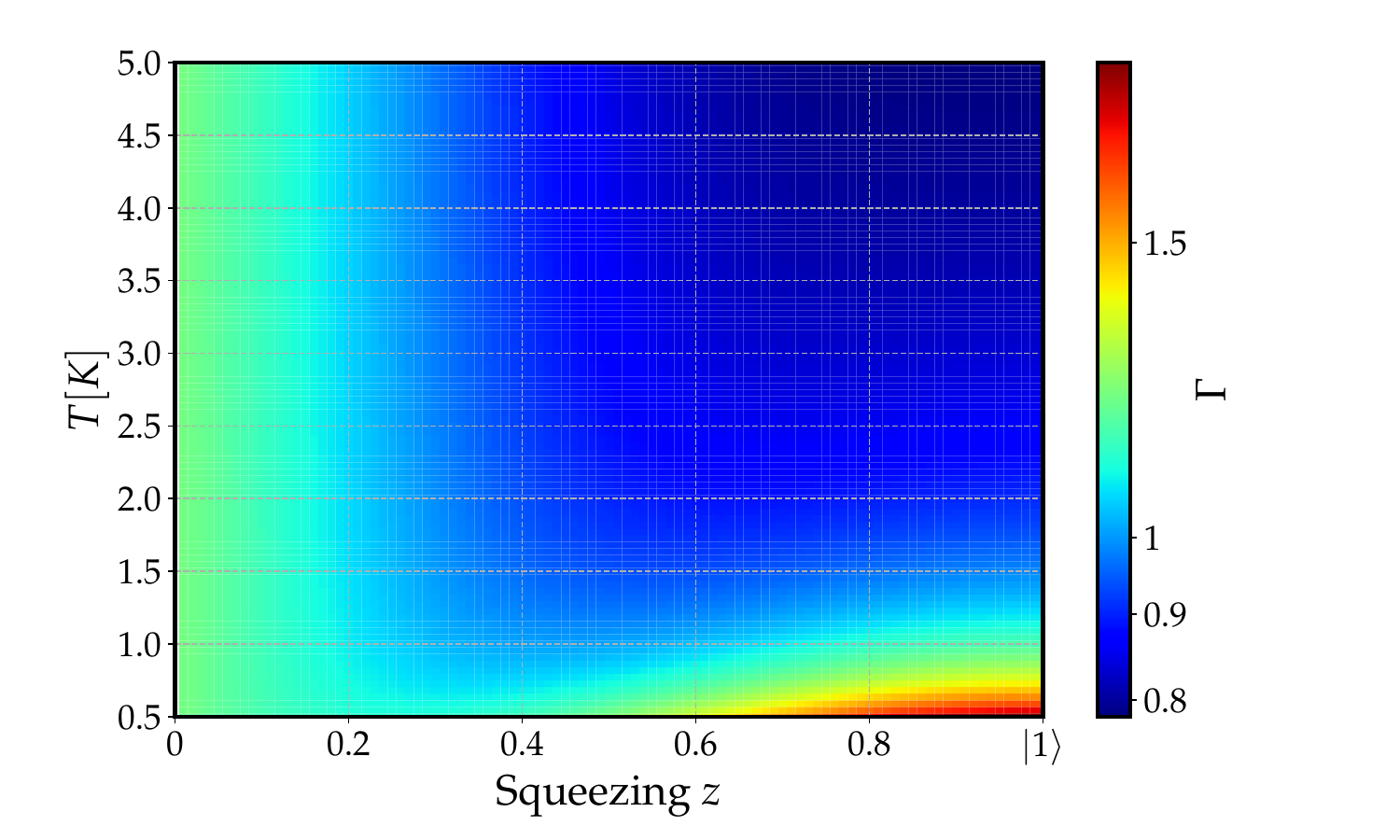}
    \caption{ \raggedright\justifying  SNR of a photon-added squeezed thermal state as a function of thermal fluctuations and squeezing. The pure Fock state $|1\rangle$ corresponds to no squeezing ($z=1$) and zero temperature. }
    \label{fock_not_optimal}
\end{figure}

At zero temperature, the simplest non-Gaussian state, namely the Fock state $|n\rangle$, has a perfectly well-defined expectation value $n$ of the photon-number observable with zero variance, therefore yielding $\Gamma\to\infty$ and becoming the optimal battery. However, as shown in Fig.~\ref{fock_not_optimal}, the situation changes dramatically in realistic scenarios. Subject to thermal fluctuations, the uncertainty in the mean photon number of thermal Fock states increases more rapidly than that of squeezed states, which are more robust to temperature increases. Consequently, a thermal regime exists where Fock states no longer provide a significant advantage in the \textit{signal-to-noise} ratio compared to the Gaussian case, and it thus becomes necessary to explore alternative protocols for precise energy extraction. 

In Ref.~\cite{Friis_2018}, Friis and Huber introduced a mathematical construction of bosonic states that minimize fluctuations and compared them to the optimal Gaussian performance within a \emph{charging} framework. More recently, Rinaldi \emph{et al.} studied charging performance using a signal-to-noise figure of merit in a Jaynes--Cummings (qubit--cavity) setting, identifying an advantage offered by pure non-classical resources \cite{rinaldi2024reliablequantumadvantagequantum}. Here we instead derive a \emph{finite-energy Gaussian bound} $\Gamma_G(\nu,\epsilon)$ for the conversion SNR $\Gamma$ (Theorem~\ref{thm:B}) under an explicit energy-input constraint $\delta N\le \epsilon$, and we exhibit \emph{physically motivated, experimentally accessible} non-Gaussian families that \emph{violate} this bound at finite temperature. Unlike Ref.~\cite{Friis_2018}, which constructs minimum-fluctuation states without an explicit detection model, and Ref.~\cite{rinaldi2024reliablequantumadvantagequantum}, which focuses on a specific Jaynes--Cummings charging setting, our analysis is tailored to \emph{bosonic batteries delivering energy to a bosonic load} and links directly to a measurable quantity (the delivered excitation number), with a simple linear photodetection model that makes the Gaussian bound directly testable from photocurrent statistics. In particular, within this framework, any observed $\Gamma>\Gamma_G(\nu,\epsilon)$ certifies that the prepared state is non-Gaussian (see Theorem~\ref{thm:B}).

Beyond the restriction to the Gaussian realm, we adopt an explicit, physically inspired ansatz:
\begin{equation}\label{eq: photon added}
    \rho_{\mathrm{NG}} \;=\; (a^\dagger)^m \,\underbrace{U_G \rho_0 U_G^\dagger}_{\rho_G}\, (a)^m, 
\end{equation}
i.e., photon addition (or subtraction) atop a Gaussian block $U_G$ (displacement and squeezing). Exact closed forms for moments via Wick’s theorem are given in \hyperref[app:non-gaussian statistics]{Appendix~C}. States engineered by \eqref{eq: photon added} arise naturally in photonic platforms via conditional operations on weakly tapped beams and single-photon detection, and multi-photon addition/subtraction is now routine \cite{roeland2022mode}. More generally, bosonic unitaries can be expanded in ladder operators, and in some cases resummed (e.g., cat/kitten states; see \hyperref[app:kitten]{Appendix~D}), so this family captures both general and practically-relevant non-Gaussian resources. Importantly, the statistical moments of \eqref{eq: photon added} can be computed \emph{exactly} by Wick’s theorem \cite{PhysRevA.96.053835}, which enables efficient evaluation of $\Gamma$ and gradient-based optimization.  All non-Gaussian expectations are computed exactly without Fock-space truncation; in \hyperref[app:truncation]{Appendix~F} we show that naive truncation can spuriously violate the Gaussian bound. Using a variational algorithm \cite{fedeypaolo}, we optimize the Gaussian block $U_G$ for fixed $m$ and constraints $(\nu,\epsilon)$.

\begin{figure}[h!]
    \centering
    \includegraphics[width=\linewidth]{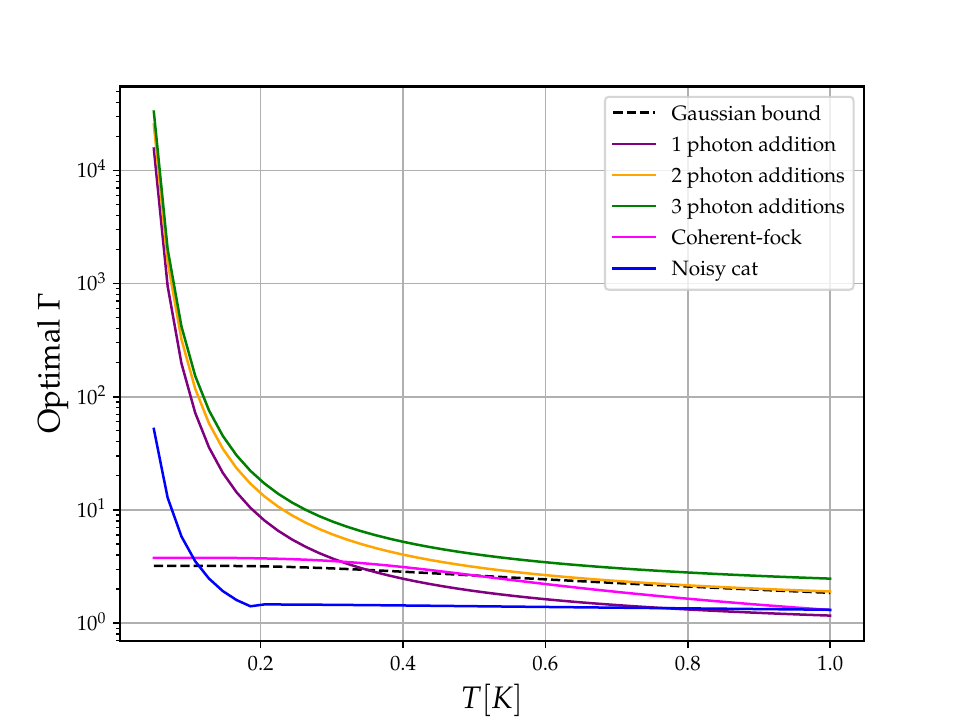}
    \caption{\raggedright\justifying Best-achieved SNR $\Gamma$ versus temperature for different non-Gaussian families (optimized over $U_G$), compared against the Gaussian bound (black dashed line). The energetic constraint has been set at $\epsilon = 5$ (dimensionless).}
    \label{snr with stellar rank}
\end{figure}

In Fig.~\ref{snr with stellar rank} we show, as a function of temperature, the highest ratio attainable for different numbers of ladder operations $m$, after optimal tuning of $U_G$. For comparison, we also plot $\Gamma$ for (i) superpositions of coherent and Fock states (with $n=5$) in contact with a thermal bath at temperature $T$ (moments from \cite{WEBER2022168986}), and (ii) \emph{noisy kitten} states modeled as photon-subtracted squeezed thermal states \cite{kitten_states,Wakui2007}. Closed-form expressions for $\Gamma$ in the kitten case are provided in \hyperref[app:kitten]{Appendix~D}.

Different non-Gaussian states \emph{violate} the Gaussian bound (black dashed line). Photon-added states recover Fock states at $T=0$, with diverging SNR as fluctuations vanish; at finite temperature, their advantage persists up to a threshold set by $(\nu,\epsilon)$ and improves with increasing $m$. Noisy cat/kitten states display a similar low-temperature advantage but are more sensitive to thermal noise than photon-added states. Conversely, coherent--Fock thermal mixtures are comparatively noise-robust, yielding stable SNRs that still surpass the Gaussian bound in the low-$\nu$ regime. The intersections of the solid curves with the Gaussian bound indicate the temperatures at which one additional photon addition is required to maintain a non-Gaussian advantage. Whether these states are feasible candidates for the charging process depends on the mean-photon constraint $\epsilon$ relevant to the task; practical operating windows and constraints are discussed in \hyperref[app:min_ergotropy_nongauss]{Appendix~E}.
We further show that, unlike the Gaussian case, passive interferometers can \emph{redistribute} moments and strictly raise $\Gamma$ for non-Gaussian resources, enabling an interferometric optimization (see Appendix Figs.~\ref{entanglement_twomodes}–\ref{fig: many_modes}).

Within the fixed-energy setting $\delta N\le\epsilon$, our results identify concrete, experimentally accessible non-Gaussian families that deliver a \emph{testable} precision advantage for energy harvesting: observing $\Gamma>\Gamma_G(\nu,\epsilon)$ is both a performance gain and, by Theorem~\ref{thm:B}, a \emph{witness} of non-Gaussianity.

\subsection{D. Multimode states} \label{subsec:multimode}
When the battery system comprises multiple modes, the possibility to perform entangling operations between them is unlocked. This naturally raises the question of whether such operations can offer an advantage in the relative precision of the stored energy \cite{Gyhm_2024}. In the most general case, each of the modes may have different oscillating frequencies, and a comment in this regard is made in \cite{Friis_2018} on the fact that Gaussian passive optics are sufficient to mix the modes in such a way that photons are stored in the lower frequency modes, implying a gain in energy precision.

However, in our scenario the response of the harvester is uniquely dependent on the photon number statistics (thus equivalent to the case where the battery modes all oscillate at the same frequency $\omega$). In that case, the following lemma applies: \\

\begin{lemma}\label{lemma1}
For multimode Gaussian states, the signal-to-noise ratio $\Gamma$ remains constant under Gaussian global (entangling) passive transformations.
\end{lemma}

The proof is provided in \hyperref[app:multimode]{Appendix~G}. On the contrary, non-Gaussian states feature \textit{signal-to-noise ratios} that are tunable by passive, entangling operations, allowing for some extra advantage with respect to the Gaussian case when the correct beam-splitting angle is applied. In this sense, our figure-of-merit also constitutes a signature of non-Gaussianity in the multimode scenario, because the dependency of $\Gamma$ on mode mixing passive interferometry is an exclusive property of non-Gaussian states. In \hyperref[app:multimode]{Appendix~G} we also discuss the scaling of the optimal $\Gamma$ with the number of modes, for Gaussian and non-Gaussian states of increasing number of photon additions, concluding that the non-Gaussian advantage is generalizable to the multi-mode scenario, and thus supporting the prospect of leveraging non-Gaussian resources for scalable and high-precision quantum batteries.

\section{IV. Implementability}  \label{secIV}
Our proposal is compatible with standard photonic hardware. The energy harvester can be a linear photodiode (Si or InGaAs PIN), whose average photocurrent is proportional to the absorbed photon flux. Writing $\eta$ for the overall quantum efficiency, $f$ for the pulse repetition rate, and $\langle N\rangle$ for the mean photons per pulse at the harvester, one has the usual relation
\begin{equation}
\langle I\rangle \;=\; q\,\eta\, f\, \langle N\rangle \;=\; R\,P_{\mathrm{opt}},
\end{equation}
with $q$ the electron charge, $P_{\mathrm{opt}}$ the mean optical power, and $R=\eta q/(h\omega)$ the responsivity \cite{RPPhotonics_Responsivity}.
In what follows we assume the standard linear photodetection regime, in which the photodiode response is well described by normally ordered photocurrent moments for a single effective optical mode within the detection bandwidth \cite{MandelWolf_OpticalCoherence,Loudon_QuantumTheoryLight}. Under this widely used model, the photocurrent statistics faithfully reproduce the photon-number statistics of the incident field up to an overall quantum efficiency $\eta$ and an additive electronic-noise contribution \cite{MandelWolf_OpticalCoherence,Loudon_QuantumTheoryLight}. All witness statements based on electrical data should be understood as conditional on these linearity and calibration assumptions, which are routinely validated in precision photodetection experiments \cite{LIGO_Squeezing_NatPhoton_2013,Berchera_2019,Ann2019}.

Shot-noise-limited readout with commercial transimpedance/balanced receivers is routine when devices are chosen for low dark current and the front-end follows standard good practice \cite{Analog_TIA_WideRange,Hamamatsu_KSPD9001E,Hamamatsu_PhotosensorModules_2025}. In this regime, with typical operation at $\langle N\rangle\sim 10^3$–$10^4$ photons per detection bin at MHz repetition rates, pulse-to-pulse fluctuations of the harvested charge are directly resolved on photodiodes. In this regime, pulsed operation enables a \emph{per-pulse electrical access} to non-classical photon-number statistics—for example, sub-shot-noise intensity fluctuations or $g^{(2)}(0)<1$—provided the linear-response assumptions above hold. Extensive prior art demonstrates that such non-classical signatures can be resolved using only photodiode readout in carefully calibrated setups \cite{LIGO_Squeezing_NatPhoton_2013,Analog_TIA_WideRange,Ann2019,Berchera_2019}. By contrast, preparing genuinely \emph{non-Gaussian} optical states at such large mean photon numbers is experimentally demanding: de-Gaussification via photon addition/subtraction is heralded and loss-sensitive, and non-Gaussian features tend to be diluted at high flux \cite{walschaers2021non,kitten_states,Wakui2007}. Accordingly, we do not rely on bright non-Gaussian sources for our benchmark.

When the optical state is too dim for single-pulse resolution—e.g., $\langle N\rangle \lesssim 10$ photons per pulse—the \emph{same} photodiode operates in an \emph{accumulate-and-reset} mode with a charge-integrating front-end: the harvester integrates the photocurrent from $M$ pulses (typically $M\!\sim\!10^4$–$10^6$, from ms to s depending on $f$) and the output voltage is sampled before reset. This replaces single-shot resolution with short-window statistics and is sufficient to observe the reduction of \emph{accumulated} fluctuations. In this operation mode one can estimate $\Gamma_{\mathrm{det}}$ from the electrical data and compare it to the \emph{detection-propagated} classical and Gaussian bounds (Sec.~V.\ref{sec:detectable_precision} and Appendix H~\ref{app:noise_detection}), thereby testing the implications of Theorems~\ref{thm:A}–\ref{thm:B} at the electrical level without modifying the optical source \cite{TI_IVC102_Integrator,Hamamatsu_KSPD9001E,Hamamatsu_PhotosensorModules_2025}. The resulting certification of non-classicality or non-Gaussianity is therefore conditional on the same linear loss/noise model used to propagate the bounds. This division of labor—bright regime for non-classicality; accumulator regime for non-Gaussianity—matches the current state of the art on non-Gaussian state preparation, where demonstrated “kitten/cat” and photon-subtracted states typically have modest mean photon numbers and are fragile to loss \cite{walschaers2021non,kitten_states,Wakui2007}.

Our theoretical analysis is frequency-agnostic (we work with a number operator $N$ that neglects spectral structure). In practice, the field has a finite spectrum and the harvester/electronics have frequency responses. Implementations therefore account for (i) the photodiode spectral responsivity $R(\omega)$ and any optical filtering, which define an effective detected mode (or set of modes), and (ii) the electronic bandwidth of the readout relative to the pulse duration and repetition rate. These standard calibrations simply enter $\eta$ and $R$ and do not change the logic of the tests (non-classicality via $g^{(2)}(0)<1$; non-Gaussianity via $\Gamma>\Gamma_G$).

Looking beyond bulk photodiodes, nanoscale transducers offer a promising outlook for ultra-low-power loads. Colloidal quantum-dot photodiodes provide high responsivity and CMOS-compatible integration across the visible–SWIR bands \cite{KonstantatosSargent_NatNano_2010}, while resonator-coupled double quantum dots have demonstrated conversion of individual \emph{microwave} photons into directed electrical current \cite{QD_DoubleDot_Harvester_PRB_2024}. Although these platforms operate in spectrally distinct bands from our optical parameter choices, they reinforce that photon-to-charge transduction at ultra-low powers is feasible with state-of-the-art technology. They also suggest that, once suitable non-Gaussian sources are interfaced to such devices, our precision-charging bounds could be tested under realistically lossy conditions.

\section{V. Detectable precision}\label{sec:detectable_precision}
Theorems~\ref{thm:A}–\ref{thm:B} are \emph{state-level} statements: $\Gamma=\delta N/\Delta N$ is a property of the photonic state delivered to the harvester and does not depend on how it is probed. In an experiment, however, we access a \emph{detection-level} SNR after a standard linear loss channel of overall efficiency $\eta$ and additive, state-independent classical readout noise. As shown in \hyperref[app:monotoneSNR]{Appendix~H.1}, such a lossy/noisy readout cannot increase SNR: the measured SNR $\Gamma_{\rm det}$ is always \emph{smaller or equal} to the state SNR $\Gamma$ (equality only in the ideal, noiseless unit-efficiency limit). To retain the witness character when testing Theorems~\ref{thm:A}–\ref{thm:B} electrically, we therefore compare $\Gamma_{\rm det}$ to the \emph{detection-propagated} classical and Gaussian bounds, obtained by pushing the corresponding state-level bounds through the same loss/noise map (see \hyperref[app:monotoneSNR]{Appendix~H.1}).

Operationally, when we say that exceeding a detection-propagated Gaussian bound “certifies non-Gaussianity at the electrical readout”, we mean the following: if (i) the optical state delivered to the harvester is the only source of non-classical fluctuations, and (ii) the harvester plus electronics are well described by a linear loss channel of efficiency $\eta$ followed by additive, state-independent classical noise, then any observed $\Gamma_{\rm det}$ larger than the detection-propagated Gaussian bound cannot be reproduced by any Gaussian input state. Under these explicit assumptions, the electrical measurement functions as an indirect yet fully task-based witness of non-Gaussianity (Theorem~\ref{thm:B}) and non-classicality (Theorem~\ref{thm:A}).

Because the harvester ultimately dissipates power in a resistive element at (effective) temperature $T_e$, the finite-time thermodynamic uncertainty relation (TUR) imposes a detection-level ceiling on any integrated-current SNR over a window $\tau$: $\Gamma_\tau \le \sqrt{\Sigma_\tau/2}$, where $\Sigma_\tau$ is the total entropy production in the load during $\tau$ \cite{BaratoSeifert2015_TUR,Gingrich2016_DissipationBounds,DechantSasa2018_EntropicBound,KoyukSeifert2019_TimeDependentTUR,HorowitzGingrich2020_NatPhys_TURReview}. We use this bound as an absolute, dissipation-limited benchmark for the electrical readout (see \hyperref[app:TUR_SM]{Appendix~H.2}); it is independent of how the optical state is prepared.

As an illustration, for $T = 1,$K, energy budget $\epsilon = 5$, overall efficiency $\eta = 0.60$, and electronic noise $\sigma_B = 0.30$, a single photon–added state optimized under these constraints attains $\Gamma_{\mathrm{det}} \approx 2.21$, while the detection-propagated Gaussian bound is $\Gamma_{G,\mathrm{det}} \approx 1.84$ (Appendix H.1).

\section{VI. Discussion and outlook}\label{sec:discussion}
We analyze the precision charging for bosonic quantum batteries under a realistic finite-energy constraint and use the \emph{signal-to-noise ratio} (SNR) of delivered excitations as an operational metric tied to what a harvester measures. On the theory side, we establish two bounds that transform this metric into \emph{witnesses}: (i) a \emph{classical} bound (Theorem~\ref{thm:A}), certifying non-classicality, and (ii) a \emph{Gaussian bound} (Theorem~\ref{thm:B}) at fixed temperature and energy input, whose violation certifies non-Gaussianity. We then identify explicit, experimentally motivated non-Gaussian families that \emph{exceed} the Gaussian bound in relevant parameter regimes and quantify how this advantage depends on temperature and energy budget. Conceptually, our results complement prior analyses of fluctuation-minimizing states in charging \cite{Friis_2018} and studies using SNR in specific models \cite{rinaldi2024reliablequantumadvantagequantum}: we provide a finite-energy Gaussian bound directly in the conversion metric measured at the load and show how concrete non-Gaussian resources surpass it.

On the implementation side, we describe a minimal, photodiode-based route that leverages standard hardware: a bright-pulse mode to access non-classical photon-number statistics per pulse on a single photodiode readout, and an accumulate-and-reset mode to test non-Gaussianity by attempting to surpass the Gaussian bound on an integrating photodiode readout. The former draws on existing sub-shot-noise demonstrations with photodiode readout. At the same time, the latter represents a realistic yet experimentally demanding next step, given the loss sensitivity and state preparation at modest mean photon numbers. Subject to standard linear-response modelling of the detector, an experimental violation of the \emph{detection-propagated} Gaussian bound at the harvester would constitute an operational quantum advantage for precision charging—namely, a performance unattainable by any classical or Gaussian state under the same energy and temperature constraints. Together, these pathways make our precision-charging witnesses testable with state-of-the-art components and realistic losses, and they point to plausible near-term use cases, such as the precision charging of fragile nanoscopic loads (e.g., photodiode or quantum-dot elements), where reducing over-threshold events at a fixed mean delivered energy is valuable.

Beyond the single-mode analysis, we explicitly consider \emph{multimode} batteries and \emph{entangling} interferometers. For \emph{Gaussian} resources, any passive linear network (beam splitters and phase shifters) that entangles the modes leaves the SNR \(\Gamma\) invariant at fixed total energy and temperature (see \hyperref[app:multimode]{Appendix~G}), whereas in the \emph{non-Gaussian} regime passive mixing redistributes moments and can \emph{strictly increase} \(\Gamma\). This yields an interferometric optimization procedure and suggests a battery-side route to \emph{entanglement witnessing} based solely on energy observables. We expect these multimode and entanglement-oriented extensions to broaden the operational scope of precision charging and to guide near-term experiments harnessing non-Gaussian resources on realistic, lossy platforms.

\textbf{Acknowledgements} \\
The authors thank Antonis Delakouras, Manuel Gessner and Francesco Flora for fruitful discussions and inputs, and notably Antonio Acín for his relevant feedback and guidance.

\textbf{Author contributions} \\
    B.P. developed the main research ideas, carried out the main calculations and simulations. F.C. proposed the main ideas and supervised the work. All authors reviewed and edited the manuscript. 
    
\textbf{Funding declaration} \\ 
This project has been funded by the European Union (EQC, 101149233), the Government of Spain (Severo Ochoa CEX2019-000910-S and FUNQIP), Fundació Cellex, Fundació Mir-Puig, Generalitat de Catalunya (CERCA program). This project is supported by the predoctoral program FI-STEP (2025 STEP 
00008), which is backed by the Secretariat of Universities and Research of the Department of Research and Universities of the Generalitat of Catalonia, as well as the European Social Plus Fund.
Views and opinions expressed are however, those of the author(s) only and do not necessarily reflect those of the European Union or European Research Executive Agency. Neither the European Union nor the granting authority can be held responsible for them.

\textbf{Competing interests} \\ 
The Authors declare no Competing Financial or Non-
Financial Interests.

\bibliography{biblio}

@article{Friis_2018,
   title={Precision and Work Fluctuations in Gaussian Battery Charging},
   volume={2},
   ISSN={2521-327X},
   url={http://dx.doi.org/10.22331/q-2018-04-23-61},
   DOI={10.22331/q-2018-04-23-61},
   journal={Quantum},
   publisher={Verein zur Forderung des Open Access Publizierens in den Quantenwissenschaften},
   author={Friis, Nicolai and Huber, Marcus},
   year={2018},
   month=apr, pages={61} }

@article{fedeypaolo,
      title={Variational quantum simulation using non-Gaussian continuous-variable systems}, 
      author={Paolo Stornati and Antonio Acin and Ulysse Chabaud and Alexandre Dauphin and Valentina Parigi and Federico Centrone},
     journal = {Phys. Rev. Res.},
  volume = {6},
  issue = {4},
  pages = {043212},
  numpages = {10},
  year = {2024},
  month = {Nov},
  publisher = {American Physical Society},
  doi = {10.1103/PhysRevResearch.6.043212},
  url = {https://link.aps.org/doi/10.1103/PhysRevResearch.6.043212}
}

@book{goldstein1995refrigerator,
  title={The refrigerator and the universe: understanding the laws of energy},
  author={Goldstein, Martin and Goldstein, Inge F},
  year={1995},
  publisher={Harvard University Press}
}

@article{binder2018thermodynamics,
  title={Thermodynamics in the quantum regime},
  author={Binder, Felix and Correa, Luis A and Gogolin, Christian and Anders, Janet and Adesso, Gerardo},
  journal={Fundamental Theories of Physics},
  volume={195},
  number={1},
  year={2018},
  publisher={Springer}
}

@article{auffeves2022quantum,
   title = {Quantum Technologies Need a Quantum Energy Initiative},
  author = {Auff\`eves, Alexia},
  journal = {PRX Quantum},
  volume = {3},
  issue = {2},
  pages = {020101},
  numpages = {12},
  year = {2022},
  month = {Jun},
  publisher = {American Physical Society},
  doi = {10.1103/PRXQuantum.3.020101},
  url = {https://link.aps.org/doi/10.1103/PRXQuantum.3.020101}
}

@misc{RPPhotonics_Responsivity,
  author       = {R{\"u}diger Paschotta},
  title        = {Responsivity (photodetectors)},
  howpublished = {RP Photonics Encyclopedia},
  year         = {2025},
  url          = {https://www.rp-photonics.com/responsivity.html},
  note         = {Accessed 24-Oct-2025}
}

@techreport{Analog_TIA_WideRange,
  title        = {Transimpedance Amplifiers for Wide Range Photodiodes Have Excellent Dynamic Performance},
  institution  = {Analog Devices},
  year         = {2021},
  url          = {https://www.analog.com/media/en/technical-documentation/technical-articles/s54_en-circuits.pdf}
}

@techreport{Hamamatsu_KSPD9001E,
  title        = {Si Photodiodes: Technical Note and Application Circuits},
  institution  = {Hamamatsu Photonics},
  year         = {2023},
  number       = {KSPD9001E},
  url          = {https://www.hamamatsu.com.cn/content/dam/hamamatsu-photonics/sites/documents/99_SALES_LIBRARY/ssd/si_pd_kspd9001e.pdf}
}

@techreport{Hamamatsu_PhotosensorModules_2025,
  title        = {Photosensor Amplifiers and Photodiode Modules},
  institution  = {Hamamatsu Photonics},
  year         = {2025},
  number       = {KACC9015E},
  url          = {https://www.hamamatsu.com/content/dam/hamamatsu-photonics/sites/documents/99_SALES_LIBRARY/ssd/photosensor_amp_kacc9015e.pdf}
}

@techreport{TI_IVC102_Integrator,
  title        = {Precision Switched Integrator Transimpedance Amplifier (IVC102 Application)},
  institution  = {Texas Instruments},
  year         = {2024},
  url          = {https://www.electronics-lab.com/wp-content/uploads/2024/03/Precision-Switched-Integrator-Transimpedance-Amplifier.pdf}
}

@article{LIGO_Squeezing_NatPhoton_2013,
  author  = {J. Aasi and {The LIGO Scientific Collaboration}},
  title   = {Enhanced sensitivity of the LIGO gravitational wave detector by using squeezed states of light},
  journal = {Nature Photonics},
  volume  = {7},
  number  = {8},
  pages   = {613--619},
  year    = {2013},
  doi     = {10.1038/nphoton.2013.177}
}

@article{QD_DoubleDot_Harvester_PRB_2024,
  title   = {Microwave power harvesting using resonator-coupled double quantum dots},
  journal = {Phys. Rev. B},
  volume  = {109},
  pages   = {L081403},
  year    = {2024},
  doi     = {10.1103/PhysRevB.109.L081403}
}

@article{KonstantatosSargent_NatNano_2010,
  author  = {G. Konstantatos and E. H. Sargent},
  title   = {Nanostructured materials for photon detection},
  journal = {Nature Nanotechnology},
  volume  = {5},
  pages   = {391--400},
  year    = {2010},
  doi     = {10.1038/nnano.2010.78}
}

@book{serafini2023quantum,
  title={Quantum continuous variables: a primer of theoretical methods},
  author={Serafini, Alessio},
  year={2023},
  publisher={CRC press}
}

@article{phillips2021energy,
  title={Energy harvesting in nanosystems: Powering the next generation of the internet of things},
  author={Phillips, Jamie D},
  journal={Frontiers in Nanotechnology},
  volume={3},
  pages={633931},
  year={2021},
  publisher={Frontiers Media SA},
url={https://doi.org/10.3389/fnano.2021.633931}
}

@article{roeland2022mode,
  title={Mode-selective single-photon addition to a multimode quantum field},
  author={Roeland, Gana{\"e}l and Kaali, Srinivasan and Rodriguez, Victor Roman and Treps, Nicolas and Parigi, Valentina},
  journal={New Journal of Physics},
  volume={24},
  number={4},
  pages={043031},
  year={2022},
  publisher={IOP Publishing},
url={http://dx.doi.org/10.1088/1367-2630/ac5f85}
}

@article{walschaers2021non,
  title={Non-Gaussian quantum states and where to find them},
  author={Walschaers, Mattia},
   journal = {PRX Quantum},
  volume = {2},
  issue = {3},
  pages = {030204},
  numpages = {68},
  year = {2021},
  month = {Sep},
  publisher = {American Physical Society},
  doi = {10.1103/PRXQuantum.2.030204},
  url = {https://link.aps.org/doi/10.1103/PRXQuantum.2.030204}}

@article{jia2024squeezing,
  title={Squeezing the quantum noise of a gravitational-wave detector below the standard quantum limit},
  author={Jia, Wenxuan and Xu, Victoria and Kuns, Kevin and Nakano, Masayuki and Barsotti, Lisa and Evans, Matthew and Mavalvala, Nergis and LIGO Scientific Collaboration and Abbott, R and Abouelfettouh, I and others},
  journal={Science},
  volume={385},
  number={6715},
  pages={1318--1321},
  year={2024},
  publisher={American Association for the Advancement of Science},
doi = {10.1126/science.ado8069},
URL = {https://www.science.org/doi/abs/10.1126/science.ado8069}
}

@article{yin2023experimental,
  title={Experimental super-Heisenberg quantum metrology with indefinite gate order},
  author={Yin, Peng and Zhao, Xiaobin and Yang, Yuxiang and Guo, Yu and Zhang, Wen-Hao and Li, Gong-Chu and Han, Yong-Jian and Liu, Bi-Heng and Xu, Jin-Shi and Chiribella, Giulio and others},
  journal={Nature Physics},
  volume={19},
  number={8},
  pages={1122--1127},
  year={2023},
  publisher={Nature Publishing Group UK London},
url={https://doi.org/10.1038/s41567-023-02046-y}
}

@article{BaratoSeifert2015_TUR,
  author  = {Barato, Andr\'e C. and Seifert, Udo},
  title   = {Thermodynamic Uncertainty Relation for Biomolecular Processes},
  journal = {Phys. Rev. Lett.},
  volume  = {114},
  number  = {15},
  pages   = {158101},
  year    = {2015},
  doi     = {10.1103/PhysRevLett.114.158101}
}

@article{Gingrich2016_DissipationBounds,
  author  = {Gingrich, Todd R. and Horowitz, Jordan M. and Perunov, Nikolai and England, Jeremy L.},
  title   = {Dissipation Bounds All Steady-State Current Fluctuations},
  journal = {Phys. Rev. Lett.},
  volume  = {116},
  number  = {12},
  pages   = {120601},
  year    = {2016},
  doi     = {10.1103/PhysRevLett.116.120601}
}

@article{DechantSasa2018_EntropicBound,
  author  = {Dechant, Andreas and Sasa, Shin-ichi},
  title   = {Current Fluctuations and Thermodynamic Uncertainty Relation in Markov Processes},
  journal = {Phys. Rev. X},
  volume  = {8},
  number  = {2},
  pages   = {021071},
  year    = {2018},
  doi     = {10.1103/PhysRevX.8.021071}
}

@article{KoyukSeifert2019_TimeDependentTUR,
  author  = {Koyuk, Tim and Seifert, Udo},
  title   = {Thermodynamic Uncertainty Relation for Time-Dependent Driving},
  journal = {Phys. Rev. Lett.},
  volume  = {122},
  number  = {23},
  pages   = {230601},
  year    = {2019},
  doi     = {10.1103/PhysRevLett.122.230601}
}

@article{HorowitzGingrich2020_NatPhys_TURReview,
  author  = {Horowitz, Jordan M. and Gingrich, Todd R.},
  title   = {Thermodynamic uncertainty relations constrain non-equilibrium fluctuations},
  journal = {Nature Physics},
  volume  = {16},
  pages   = {15--20},
  year    = {2020},
  doi     = {10.1038/s41567-019-0702-6}
}

@article{mishra2012review,
  title={A review on supply of power to Nanorobots used in nanomedicine},
  author={Mishra, Kailash Chandra},
  journal={International Journal of Advances in Engineering \& Technology},
  volume={4},
  number={2},
  pages={564},
  year={2012},
  publisher={IAET Publishing Company},
url={https://www.ijaet.org/media/0007/63I10-IJAET1009228-A-REVIEW-ON-SUPPLY.pdf}
}

@article{ussia2024unlocking,
  title={Unlocking the potential and versatility of quantum dots: from biomedical to environmental applications and smart micro/nanorobots},
  author={Ussia, Martina and Privitera, Vittorio and Scalese, Silvia},
  journal={Advanced Materials Interfaces},
  volume={11},
  number={17},
  pages={2300970},
  year={2024},
  publisher={Wiley Online Library},
url={ https://doi.org/10.1002/admi.202300970}
}

@article{lahaye2004approaching,
  title={Approaching the quantum limit of a nanomechanical resonator},
  author={LaHaye, MD and Buu, Olivier and Camarota, Benedetta and Schwab, KC},
  journal={Science},
  volume={304},
  number={5667},
  pages={74--77},
  year={2004},
  publisher={American Association for the Advancement of Science},
doi = {10.1126/science.1094419},
URL = {https://www.science.org/doi/abs/10.1126/science.1094419}
}

@article{zhou2021magnetically,
  title={Magnetically driven micro and nanorobots},
  author={Zhou, Huaijuan and Mayorga-Martinez, Carmen C and Pan{\'e}, Salvador and Zhang, Li and Pumera, Martin},
  journal={Chemical Reviews},
  volume={121},
  number={8},
  pages={4999--5041},
  year={2021},
  publisher={ACS Publications},
url={https://doi.org/10.1021/acs.chemrev.0c01234}
}

@article{giri2021brief,
  title={A brief review on challenges in design and development of nanorobots for medical applications},
  author={Giri, Gautham and Maddahi, Yaser and Zareinia, Kourosh},
  journal={Applied Sciences},
  volume={11},
  number={21},
  pages={10385},
  year={2021},
  publisher={MDPI},
URL = {https://www.mdpi.com/2076-3417/11/21/10385}
}

@article{myers2022quantum,
 title={Quantum thermodynamic devices: From theoretical proposals to experimental reality},
   volume={4},
   ISSN={2639-0213},
   url={http://dx.doi.org/10.1116/5.0083192},
   DOI={10.1116/5.0083192},
   number={2},
   journal={AVS Quantum Science},
   publisher={American Vacuum Society},
   author={Myers, Nathan M. and Abah, Obinna and Deffner, Sebastian},
   year={2022},
   month=apr }

@article{culhane2022extractable,
  title={Extractable work in quantum electromechanics},
  author={Culhane, Ois{\'\i}n and Mitchison, Mark T and Goold, John},
  journal={Physical Review E},
  volume={106},
  number={3},
  pages={L032104},
  year={2022},
  publisher={APS},
doi = {10.1103/PhysRevE.106.L032104},
  url = {https://link.aps.org/doi/10.1103/PhysRevE.106.L032104}
}

@article{jaliel2019experimental,
  title={Experimental realization of a quantum dot energy harvester},
  author={Jaliel, Gulzat and Puddy, RK and S{\'a}nchez, R and Jordan, AN and Sothmann, B and Farrer, I and Griffiths, JP and Ritchie, DA and Smith, CG},
  journal={Physical review letters},
  volume={123},
  number={11},
  pages={117701},
  year={2019},
  publisher={APS},
 doi = {10.1103/PhysRevLett.123.117701},
  url = {https://link.aps.org/doi/10.1103/PhysRevLett.123.117701}
}

@article{meng2024quantum,
  title={Quantum harvester enables energy transfer without randomness transfer or dissipation},
  author={Meng, Fei and Xu, Junhao and Liu, Xiangjing and Dahlsten, Oscar},
  journal = {Phys. Rev. A},
  volume = {111},
  issue = {1},
  pages = {012219},
  numpages = {16},
  year = {2025},
  month = {Jan},
publisher = {American Physical Society},
  doi = {10.1103/PhysRevA.111.012219},
  url = {https://link.aps.org/doi/10.1103/PhysRevA.111.012219}
}

@article{campaioli2024colloquium,
  title = {Colloquium: Quantum batteries},
  author = {Campaioli, Francesco and Gherardini, Stefano and Quach, James Q. and Polini, Marco and Andolina, Gian Marcello},
  journal = {Rev. Mod. Phys.},
  volume = {96},
  issue = {3},
  pages = {031001},
  numpages = {30},
  year = {2024},
  month = {Jul},
  publisher = {American Physical Society},
  doi = {10.1103/RevModPhys.96.031001},
  url = {https://link.aps.org/doi/10.1103/RevModPhys.96.031001}
}

@article{PhysRev.130.2529,
  title = {The Quantum Theory of Optical Coherence},
  author = {Glauber, Roy J.},
  journal = {Phys. Rev.},
  volume = {130},
  issue = {6},
  pages = {2529--2539},
  numpages = {0},
  year = {1963},
  month = {Jun},
  publisher = {American Physical Society},
  doi = {10.1103/PhysRev.130.2529},
  url = {https://link.aps.org/doi/10.1103/PhysRev.130.2529}
}

@book{MandelWolf_OpticalCoherence,
  author    = {L. Mandel and E. Wolf},
  title     = {Optical Coherence and Quantum Optics},
  publisher = {Cambridge University Press},
  year      = {1995}
}

@book{Loudon_QuantumTheoryLight,
  author    = {R. Loudon},
  title     = {The Quantum Theory of Light},
  edition   = {3},
  publisher = {Oxford University Press},
  year      = {2000}
}

@article{TAN2019100030,
title = {The resurgence of the linear optics quantum interferometer — recent advances \& applications},
author = {Si-Hui Tan and Peter P. Rohde},
journal = {Reviews in Physics},
volume = {4},
pages = {100030},
year = {2019},
issn = {2405-4283},
url = {https://www.sciencedirect.com/science/article/pii/S2405428318300431}
}

@article{PhysRevA.96.053835,
  title = {Statistical signatures of multimode single-photon-added and -subtracted states of light},
  author = {Walschaers, Mattia and Fabre, Claude and Parigi, Valentina and Treps, Nicolas},
  journal = {Phys. Rev. A},
  volume = {96},
  issue = {5},
  pages = {053835},
  numpages = {23},
  year = {2017},
  month = {Nov},
  publisher = {American Physical Society},
  doi = {10.1103/PhysRevA.96.053835},
  url = {https://link.aps.org/doi/10.1103/PhysRevA.96.053835}
}

@article{WEBER2022168986,
title = {Non-classical properties of superposition thermal quantum states},
journal = {Annals of Physics},
volume = {443},
pages = {168986},
year = {2022},
issn = {0003-4916},
doi = {https://doi.org/10.1016/j.aop.2022.168986},
url = {https://www.sciencedirect.com/science/article/pii/S0003491622001415},
author = {P.E.R. Weber and V.N.A. Lula-Rocha and J.C.C. Pereira and M.A.S. Trindade and L.M. Silva Filho and M.G.R. Martins and A.E. Santana and J.D.M. Vianna}
}

@article{kitten_states,
author = {Alexei Ourjoumtsev  and Rosa Tualle-Brouri  and Julien Laurat  and Philippe Grangier },
title = {Generating Optical Schrödinger Kittens for Quantum Information Processing},
journal = {Science},
volume = {312},
number = {5770},
pages = {83-86},
year = {2006},
URL = {https://www.science.org/doi/abs/10.1126/science.1122858}}

@article{Wakui2007,
  title = {Photon subtracted squeezed states generated with periodically poled $\text{KTiOP}\text{O}_4$},
author = {Wakui,  Kentaro and Takahashi,  Hiroki and Furusawa,  Akira and Sasaki,  Masahide},
  volume = {15},
  ISSN = {1094-4087},
  number = {6},
  journal = {Optics Express},
  publisher = {Optica Publishing Group},
  year = {2007},
  URL = {http://dx.doi.org/10.1364/OE.15.003568},
  pages = {3568}
}

@article{QUACH20232195,
title = {Quantum batteries: The future of energy storage?},
journal = {Joule},
volume = {7},
number = {10},
pages = {2195-2200},
year = {2023},
issn = {2542-4351},
doi = {https://doi.org/10.1016/j.joule.2023.09.003},
url = {https://www.sciencedirect.com/science/article/pii/S2542435123003641},
author = {J.Q. Quach and G. Cerullo and T. Virgili}
}

@article{Camposeo_2025,
   title={Quantum Batteries: A Materials Science Perspective},
   volume={37},
   ISSN={1521-4095},
   url={http://dx.doi.org/10.1002/adma.202415073},
   DOI={10.1002/adma.202415073},
   number={17},
   journal={Advanced Materials},
   publisher={Wiley},
   author={Camposeo, Andrea and Virgili, Tersilla and Lombardi, Floriana and Cerullo, Giulio and Pisignano, Dario and Polini, Marco},
   year={2025},
   month=feb }

@misc{shi2025quantumchargingadvantagemultipartite,
  author = {Shi, Hai-Long and Gan, Li and Zhang, Kun and Wang, Xiao-Hui and Yang, Wen-Li},
  title  = {Quantum Charging Advantage from Multipartite Entanglement},
  year   = {2025},
  eprint = {arXiv:2503.02667},
  note   = {arXiv:2503.02667 [quant-ph]},
  url    = {https://arxiv.org/abs/2503.02667}
}

@article{Konar_2024,
   title={Multimode advantage in continuous-variable quantum batteries},
   volume={110},
   ISSN={2469-9934},
   url={http://dx.doi.org/10.1103/PhysRevA.110.022226},
   DOI={10.1103/physreva.110.022226},
   number={2},
   journal={Physical Review A},
   publisher={American Physical Society (APS)},
   author={Konar, Tanoy Kanti and Patra, Ayan and Gupta, Rivu and Ghosh, Srijon and Sen(De), Aditi},
   year={2024},
   month=aug }

@misc{kurman2025quantumcomputationquantumbatteries,
      title={Quantum Computation with Quantum Batteries}, 
      author={Yaniv Kurman and Kieran Hymas and Arkady Fedorov and William J. Munro and James Quach},
      year={2025},
      eprint={2503.23610},
      archivePrefix={arXiv},
      primaryClass={quant-ph},
      url={https://arxiv.org/abs/2503.23610}, 
}

@article{Liu2021,
  title = {Experimental critical quantum metrology with the Heisenberg scaling},
  volume = {7},
  ISSN = {2056-6387},
  url = {http://dx.doi.org/10.1038/s41534-021-00507-x},
  DOI = {10.1038/s41534-021-00507-x},
  number = {1},
  journal = {npj Quantum Information},
  publisher = {Springer Science and Business Media LLC},
  author = {Liu,  Ran and Chen,  Yu and Jiang,  Min and Yang,  Xiaodong and Wu,  Ze and Li,  Yuchen and Yuan,  Haidong and Peng,  Xinhua and Du,  Jiangfeng},
  year = {2021},
  month = dec
}

@article{PhysRevA.90.063824,
  title = {Antibunching and unconventional photon blockade with Gaussian squeezed states},
  author = {Lemonde, Marc-Antoine and Didier, Nicolas and Clerk, Aashish A.},
  journal = {Phys. Rev. A},
  volume = {90},
  issue = {6},
  pages = {063824},
  numpages = {11},
  year = {2014},
  month = {Dec},
  publisher = {American Physical Society},
  doi = {10.1103/PhysRevA.90.063824},
  url = {https://link.aps.org/doi/10.1103/PhysRevA.90.063824}
}

@article{Berchera_2019,
   title={Quantum imaging with sub-Poissonian light: challenges and perspectives in optical metrology},
   volume={56},
   ISSN={1681-7575},
   url={http://dx.doi.org/10.1088/1681-7575/aaf7b2},
   DOI={10.1088/1681-7575/aaf7b2},
   number={2},
   journal={Metrologia},
   publisher={IOP Publishing},
   author={Berchera, I Ruo and Degiovanni, I P},
   year={2019},
   month=jan, pages={024001} }

@article{Ann2019,
  title = {Observation of scalable sub-Poissonian-field lasing in a microlaser},
  volume = {9},
  ISSN = {2045-2322},
  url = {http://dx.doi.org/10.1038/s41598-019-53525-3},
  DOI = {10.1038/s41598-019-53525-3},
  number = {1},
  journal = {Scientific Reports},
  publisher = {Springer Science and Business Media LLC},
  author = {Ann,  Byoung-moo and Song,  Younghoon and Kim,  Junki and Yang,  Daeho and An,  Kyungwon},
  year = {2019},
  month = nov 
}

@misc{brask2022gaussianstatesoperations,
      title={Gaussian states and operations -- a quick reference}, 
      author={Jonatan Bohr Brask},
      year={2022},
      eprint={2102.05748},
      archivePrefix={arXiv},
      primaryClass={quant-ph},
      url={https://arxiv.org/abs/2102.05748}, 
}

@article{Zhang2022,
  title = {Universal interference-based construction of Gaussian operations in hybrid quantum systems},
  volume = {8},
  ISSN = {2056-6387},
  url = {http://dx.doi.org/10.1038/s41534-022-00581-9},
  DOI = {10.1038/s41534-022-00581-9},
  number = {1},
  journal = {npj Quantum Information},
  publisher = {Springer Science and Business Media LLC},
  author = {Zhang,  Mengzhen and Chowdhury,  Shoumik and Jiang,  Liang},
  year = {2022},
  month = jun 
}

@article{loophaffnian,
  title={A Faster Hafnian Formula for Complex Matrices and Its Benchmarking on a Supercomputer},
  author={Andreas Bj{\"o}rklund and Brajesh Gupt and Nicol{\'a}s Quesada},
  journal={Journal of Experimental Algorithmics (JEA)},
  year={2018},
  volume={24},
  pages={1 - 17},
  url={https://api.semanticscholar.org/CorpusID:141413702}
}

@misc{rinaldi2024reliablequantumadvantagequantum,
      title={Reliable quantum advantage in quantum battery charging}, 
      author={Davide Rinaldi and Radim Filip and Dario Gerace and Giacomo Guarnieri},
      year={2024},
      eprint={2412.15339},
      archivePrefix={arXiv},
      primaryClass={quant-ph},
      url={https://arxiv.org/abs/2412.15339}, 
}

@article{Gyhm_2024,
   title={Beneficial and detrimental entanglement for quantum battery charging},
   volume={6},
   ISSN={2639-0213},
   url={http://dx.doi.org/10.1116/5.0184903},
   DOI={10.1116/5.0184903},
   number={1},
   journal={AVS Quantum Science},
   publisher={American Vacuum Society},
   author={Gyhm, Ju-Yeon and Fischer, Uwe R.},
   year={2024},
   month=jan }
\clearpage
\onecolumngrid

\appendix

\section{Appendix A: Covariance matrix formalism for Gaussian states}\label{app:covmatrix}
Gaussian states can be understood as the set of all ground and thermal states of Hamiltonians that are at most quadratic in the quadrature variables $x$ and $p$. These quadratures, also referred to as canonical operators, can be chosen to be any pair of operators acting on the system's Hilbert space and satisfying the canonical commutation relations:
\begin{equation}
    [x_i, p_j] = i \hbar \delta_{ij}
\end{equation}
In the harmonic oscillator picture, they are usually represented by the position and momentum of a particle, but when dealing with quantum states of light, it is the real and imaginary components of the electromagnetic field that play the roles of the quadrature variables. \\

This Gaussian set entails the operational advantage that all of its states have positively-valued normal distributions in phase space, and thus admit a full characterization through the first and second statistical moments of $x$ and $p$, which are encoded in a vector of displacements, and the so-called covariance matrix, respectively. Knowledge of these two degrees of freedom is thus sufficient to extract any information of interest from a Gaussian state. \\

There exist various possibilities for the parametrization of a Gaussian state's covariance matrix. For the sake of a greater intuition on the properties of the states on which we are computing the \textit{signal-to-noise} ratio $\Gamma$, we will adopt the typical approach used in the context of quantum optics and photonics that describes such matrix in terms of the transformations applied to an initially thermal state in an optical setup. According to Williamson's theorem, the covariance matrix of any Gaussian state can be written as the product:
\begin{equation}
    \sigma = S \bigg( \bigoplus_{j=1}^N 
\nu_j \Id_2 \bigg) S^T
\end{equation}
where $S \in Sp_{2n,\mathds{R}}$ is a symplectic transformation that can also be, through the Bloch-Messiah decomposition theorem, split into:
\begin{equation}
    S = O_1 Z O_2
\end{equation}
with
    \begin{equation*}
O_1,O_2 \in O(2N) \cap Sp_{2n,\mathds{R}} \cong U(N) 
\end{equation*}
representing the set of passive optics unitaries, and
\begin{equation}
    Z= \bigoplus_{j=1}^N\begin{pmatrix}
z_j & 0 \\
0 & z_j^{-1} \end{pmatrix} \quad z \in (0,1] 
\end{equation}
a squeezing transformation of parameter $z$.
This representation allows interpreting every Gaussian state as the result of applying a single-mode squeezing transformation, followed by a linear-optical passive setup, to an initial thermal (and possibly displaced) state of $N$ modes. Such initial state would have covariance matrix $\sigma_0 = \bigoplus_{j=1}^N 
\nu_j \Id_2$, where the $\nu_i$ are referred to as symplectic eigenvalues, and can be understood in this context as fluctuation parameters indicating the temperature $T_i$ that characterizes the initial thermal state of each of the modes through the relation: 
\begin{equation}
    \nu_i = \coth \bigg(\frac{\omega_i}{k_B \ T_i}\bigg)
\end{equation}

The matrices $O_1, O_2$ represent the energy-preserving  \textit{passive optics} transformations, which can be decomposed into beam splitters (semi-reflective mirrors parametrized by their transmissivity $\cos^2\theta$ that mix up pairs of modes) and phase shifters (dielectric plates that rotate the optical phase of an electromagnetic wave by some angle $\varphi$). A scheme for alternatingly applying these two operations to construct the most general linear interferometer is detailed in \cite{TAN2019100030}. Matrix $Z$ represents local or single-mode squeezing, which, in contrast, is an active operation, requiring energy from an external laser field that pumps a nonlinear crystal through which the optical modes travel. The result of this transformation is a contraction of the variance of one of the canonical variables and the expansion of its conjugate one. Squeezing and beam splitting enable the generation of quantum correlations between different modes of a quantum state. 

The most general single-mode Gaussian covariance matrix in terms of the elementary quantum optical operations takes the form: 
\begin{equation}\label{cov_gaussian}
    \sigma = \nu \begin{pmatrix}
        z \cos^2 \varphi + \frac{1}{z} \sin^2 \varphi & \frac{1-z^2}{2z} \sin(2\varphi) \\ \\  \frac{1-z^2}{2z} \sin(2\varphi) & z \sin^2 \varphi + \frac{1}{z} \cos^2 \varphi
    \end{pmatrix}
\end{equation}
The mean and variance of the photon number operator can be obtained through (see \cite{Friis_2018})
\begin{equation} \label{expvalN_gaussian}
    N(\rho) = \frac{1}{4} (\operatorname{Tr}(\sigma) - 2) +  \lVert \bm{\alpha}\rVert ^2
\end{equation}
\begin{equation} \label{deltaN_gaussian}
    \Delta N (\rho)=  \sqrt{\bm{\alpha}^T \sigma \bm{\alpha}  + \frac{1}{8}[\operatorname{Tr}(\sigma^2) -2 ]}
\end{equation}
and thus depend, upon substitution of $\sigma$ as in Eq.\ref{cov_gaussian},  on the fluctuation parameter $\nu$ that accounts for temperature, the squeezing factor $z \in (0,1]$, the dephasing angle $\varphi \in [0,2\pi)$ and the displacement vector $\bm{\alpha}$:

\begin{equation}\label{ergotropy_gaussian}
    N (\rho) -  N (\rho_0) = \frac{1}{4} \nu \bigg(z + \frac{1}{z} -2 \bigg) + \lVert \bm{\alpha} \rVert ^2
\end{equation}
\begin{align}\label{eq: long_N2}
\begin{split}
    (\Delta N(\rho))^2 = \frac{1}{8} \nu^2 \bigg(z^2 + \frac{1}{z^2} \bigg) -\frac{1}{4}  +  \nu \bigg[ \alpha_1^2 (z \cos^2 \varphi + \frac{1}{z} \sin^2 \varphi) +  \alpha_2^2 (z \sin^2 \varphi + \frac{1}{z} \cos^2 \varphi)  + 2 \alpha_1 \alpha_2 \bigg(\frac{1-z^2}{z} \sin\varphi \  \cos\varphi \bigg)\bigg]
\end{split}
\end{align}
Eq.\ref{eq: long_N2} can be simplified by noting that the phase shift angle $\varphi$ can be absorbed without loss of generality in the vector of displacements through a rotation $R(\varphi)$:
\begin{equation}
\begin{pmatrix}
        \alpha_1 \\  \alpha_2 \end{pmatrix} = \begin{pmatrix}
         \cos(\varphi) &  \sin(\varphi) \\   -\sin(\varphi) &  \cos (\varphi)
    \end{pmatrix} \begin{pmatrix}
       \alpha_1' \\  \alpha_2'
    \end{pmatrix}
\end{equation}
and so the uncertainty in photon number yields
\begin{equation} \label{deltaN_simplified}
    \Delta N = \sqrt{ \frac{1}{8} \nu^2 \bigg(z^2 + \frac{1}{z^2} \bigg) -\frac{1}{4}  +  \nu \bigg(\alpha_1^2 z+\alpha_2^2 \frac{1}{z}\bigg)}
\end{equation}
and through the quotient between Eqs.\ref{ergotropy_gaussian} and \ref{deltaN_simplified} one can compute the ratio $\Gamma$ of any Gaussian state as a function of the squeezing and displacement transformations.\\
\section{Appendix B: Optimal \textit{signal-to-noise} ratio $\Gamma$ for Gaussian states} \label{app: optimal_gaussian}
In order to obtain the Gaussian transformation that minimizes the uncertainty in the harvested energy, constrained to a fixed temperature of the system and to some maximum amount of such energy, we apply the method of Lagrange multipliers to the following optimization problem:
\begin{align} \label{optimization}
\begin{split}
\textit{Minimize} & \\ f(z,\alpha_1,\alpha_2)  & =  \frac{1}{8} \nu^2 \bigg(z^2 + \frac{1}{z^2} \bigg)  +  \nu \bigg(\alpha_1^2 z+\alpha_2^2 \frac{1}{z}\bigg) \\
\textit{Subject to} & \\ g(z,\alpha_1,\alpha_2)  & =  \frac{1}{4} \nu \bigg(z + \frac{1}{z} -2 \bigg) + |\alpha|^2 \leq \epsilon  \\ & 0  < z \leq 1
\end{split}
\end{align}
This strategy first provides us with the optimal Gaussian squeezing $z$ (as a function of the two fixed parameters) in the form of an implicit quartic equation. We are interested in the solution that is a physical value of squeezing (i.e., is real and lies within the interval $(0,1]$).
\begin{equation} \label{eq: optimal_squeezing}
    \nu \bigg(\frac{1}{z^3_{\text{opt}}}+z_{\text{opt}}-2 \bigg) = 4\epsilon \quad \quad z \in \mathbb{R}, z\in (0,1]
\end{equation}
According to Lagrange's equations, the optimal displacement $\Vec{\alpha}$ parameters depend on the previously obtained optimal squeezing through:
\begin{align}
    \begin{split}
        \alpha_{1,o p t}^2 &= \frac{\nu}{4} \bigg(\frac{1}{z^3_{\text{opt}}} - \frac{1}{z_{\text{opt}}} \bigg) \\
        \alpha_{2,o p t}^2 &= 0 
    \end{split}
\end{align}
and finally the optimal Gaussian ratios $\Gamma$, which are shown in Fig.2 in the main text,  are simply calculated by substituting these optimal solutions for $z$ and $\Vec{\alpha}$ into Eqs.\ref{ergotropy_gaussian} and \ref{deltaN_simplified}.\\

\begin{figure}[h!]
    \centering
    \includegraphics[width=0.7\linewidth]{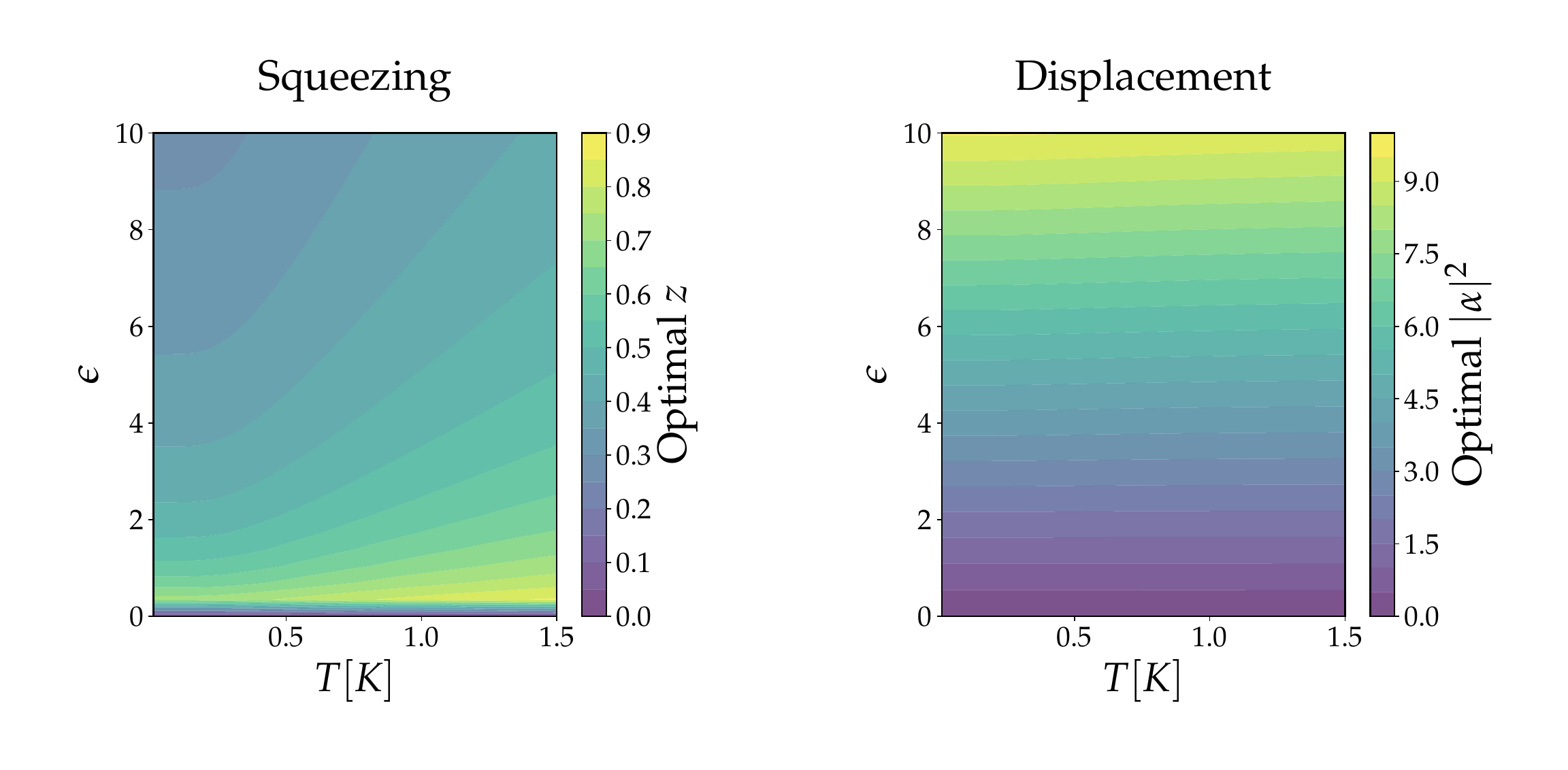}
    \caption{\raggedright\justifying Solution to \ref{optimization} for the optimal squeezing and displacement through the Lagrange multipliers method.}
    \label{lagrange_opt_parameters}
\end{figure}

To get a hint on how much of each energetic resource (squeezing and displacement) contributes to the optimal Gaussian state, we show the numerical values of the solutions as a function of $T, \epsilon$ in Fig.\ref{lagrange_opt_parameters}. Whereas the optimal displacement appears to increase with increasing energy input to the system, regardless of temperature, the optimal squeezing shows a less uniform behaviour. Interestingly, for most of the $(T, \epsilon)$ values, highly squeezed states (i.e. $z_{o p t}$ small) are advantageous, and so for the majority of the parameter space we may approximate  
\begin{equation}
    \frac{1}{z^3_{\text{opt}}} \gg z_{\text{opt}}
\end{equation} 
and solve Eq.\ref{eq: optimal_squeezing} explicitly for the squeezing parameter:
\begin{equation}
    z_{\text{opt}} \approx \bigg(\frac{4\epsilon}{\nu} +2 \bigg)^{-\frac{1}{3}}
\end{equation}

 With this approximation, the optimal Gaussian \textit{signal-to-noise} ratio finally yields:

\begin{equation}
    \Gamma_{o p t} \approx \frac{8\epsilon}{\nu^2\bigg[3\bigg(\frac{4\epsilon}{\nu}+2\bigg)^{\frac{2}{3}}-2+\bigg(\frac{4\epsilon}{\nu}+2\bigg)^{\frac{-2}{3}} \bigg]}
\end{equation}

\section{Appendix C: Expectation values on non-Gaussian states} \label{app:non-gaussian statistics}
The procedure for computing the expectation value and uncertainty of the photon number operator to optimize our objective quantity $\Gamma$ assumes that any quantum operator $M$ can be written as a product of ladder operators, and also that the output state after the whole circuit can be written as another product of ladder operators acting on an underlying Gaussian state with density matrix $\rho_G$ :
$$
M=\prod_{j=1}^{n} a_j^{\#}, \quad \rho_{N G}=\frac{1}{K} \prod_{i=1}^{m} a_i^{\#} \rho_G (a_i^{\#})^\dagger
$$
where $\# \in\{\dagger, \cdot\}$ denotes the type of ladder operator (creation or annihilation, respectively), and $K:=\operatorname{Tr}\left[ \prod_{i=1}^{m} a_i^{\#} \rho_G (a_i^{\#})^\dagger\right]$ is a normalization factor that comes from the fact that ladder operators are non-unitary. By the cyclic property of the trace:

\begin{equation}
\operatorname{Tr}\left[M\rho_{NG}\right] = \frac{1}{K}\operatorname{Tr}\left[a_{S_m}^{\dagger} \ldots\right. \left.a_{S_1}^{\dagger} a_{C_n}^{\#} \ldots a_{C_1}^{\#}  a_{S_1} \ldots a_{S_m} \rho_G\right] 
\end{equation}
where operators with subscript $S_i$ are those implementing the photon subtractions (analogous for additions by taking hermitian conjugates) on mode $i$ of the Gaussian state, and those with subscript $C_j$ comprise the expression of the observable $M$ whose expectation value we wish to calculate. From Wick's theorem, we know that the expectation value of a single product of ladder operators can be decomposed as the following sum: 
\begin{equation}
 \label{trace}
\operatorname{Tr} \left[a_{S_m}^{\dagger} \ldots\right. \left.a_{S_1}^{\dagger} a_{C_n}^{\#} \ldots a_{C_1}^{\#}  a_{S_1} \ldots a_{S_m} \rho_G\right] = \sum_{\mathcal{P}} \prod_{\left\{\left(p_1, \#\right),\left(p_2, \#\right)\right\} \in \mathcal{P}} \operatorname{Tr}\left[a_{p_1}^{\#} a_{p_2}^{\#} \rho_G\right]
 \end{equation}
where $\mathcal{P}$ is the set of all possible perfect matchings of the $2m+n$ indices $\left(p_k, \#\right) \in\left\{C_1, \ldots, C_n, S_1, \ldots, S_m\right\} \times$ $\{\dagger, \cdot\}$ of the ladder operators, including loops (i.e. matchings of one of the operators with itself \cite{loophaffnian}). All we are left with are expectation values of singlets or pairs of annihilation or creation operators on a Gaussian state, which can be computed by using the following identities:
\begin{equation} \label{identities}
\begin{aligned}
d_1 & = \operatorname{Tr}\left[a_j  \rho_G\right] = \alpha_j \\
d_2 & = 
 \operatorname{Tr}\left[a_j^{\dagger}  \rho_G\right] = \alpha_j^* \\
I_1 & =\operatorname{Tr}\left[a_j^{\dagger} a_k^{\dagger} \rho_G\right] =\frac{1}{4}\left[V_{j k}-V_{j+N, k+N}-i\left(V_{j, k+N}+V_{j+N, k}\right)\right] \\
I_2 & =\operatorname{Tr}\left[a_j a_k \rho_G\right]=I_1^* \\
I_3 & =\operatorname{Tr}\left[a_j^{\dagger} a_k \rho_G\right] =\frac{1}{4}\left[V_{j k}+V_{j+N, k+N}+i\left(V_{j, k+N}-V_{j+N, k}\right)-2 \delta_{j k}\right]\\
I_4 & =\operatorname{Tr}\left[a_j a_k^{\dagger} \rho_G\right]=\delta_{j k}+I_3^{*}
\end{aligned}
\end{equation}

where $\alpha_j$ and $V_{j k}$ are the j-th mode's diplacement and the $(j, k)$ matrix element of the $2 N \times 2 N$ covariance matrix of the quadratures of the state $\rho_G$, respectively. 

\section{Appendix D: Kitten states example}\label{app:kitten}
Cat states, defined in quantum optics as superpositions of coherent states with opposite phases, present significant challenges for experimental realization. However, it has been shown that these states can be approximated--hence the term \textit{kitten}--by successively applying photon subtractions to a squeezed thermal state \cite{kitten_states,Wakui2007} (cat states would correspond to the limit of infinite subtractions). This approach not only simplifies their practical implementation but also yields a particular class of non-Gaussian distributions for which the statistical moments of the photon-number observable can be handled analytically. \\

Let us consider the covariance matrix of a thermal squeezed state of a single mode
$$ \sigma = \bigg(\begin{array}{cc}
   \nu z  & 0 \\
   0  & \frac{\nu}{z}
\end{array} \bigg)$$
where $\nu$ is the noise parameter and $z$ accounts or squeezing. Since we are not considering displacement, $\bm{\alpha} = \bm{0}$. The Wick identities (Eq.\ref{identities}) for the corresponding state take the form:
\begin{equation}\label{identities_cat}
    \begin{aligned}
        d_1 & = d_2 =0 \\
        I_1 & = I_2 = \frac{\nu}{4} \bigg(z - \frac{1}{z}\bigg) \\
        I_3 & = \frac{\nu}{4} \bigg(z + \frac{1}{z}\bigg) - \frac{1}{2} \\
        I_4 &  = \frac{\nu}{4} \bigg(z + \frac{1}{z}\bigg) + \frac{1}{2}  
    \end{aligned}
\end{equation}
Now, suppose we perform $m$ successive photon subtractions on the previous thermal squeezed state. The mean values of the photon number and its square for the output non-Gaussian state read:
\begin{equation} \label{n_cat}
    N_{c a t} = \frac{1}{K}  \operatorname{Tr}[\ \underbrace{a^\dagger a}_N \ \underbrace{a \ldots a}_m \ \sigma \ \underbrace{a^\dagger \ldots a^\dagger}_m ] \ = \quad \frac{1}{K}  \operatorname{Tr}[\ \underbrace{a^\dagger \ldots a^\dagger}_{m+1} \ \underbrace{a \ldots a}_{m+1} \ \sigma  ]  
\end{equation}
\begin{equation} \label{n2_cat}
\begin{aligned}
    N_{c a t}^2 &= \frac{1}{K}  \operatorname{Tr}[\ \underbrace{a^\dagger a a^\dagger a}_{N^2} \ \underbrace{a \ldots a}_m \ \sigma \ \underbrace{a^\dagger \ldots a^\dagger}_m ] \ = \quad \frac{1}{K}  \operatorname{Tr}[\ \underbrace{a^\dagger \ldots a^\dagger}_{m+1} \ a a^\dagger \ \underbrace{a \ldots a}_{m+1} \ \sigma  ] \  \stackrel{[a , a^\dagger]=1 }{=} \\
    &  = \frac{1}{K} \bigg(  \operatorname{Tr}[\ \underbrace{a^\dagger \ldots a^\dagger}_{m+1} \ \underbrace{a \ldots a}_{m+1} \ \sigma  ] +  \operatorname{Tr}[\ \underbrace{a^\dagger \ldots a^\dagger}_{m+2}  \ \underbrace{a \ldots a}_{m+2} \ \sigma  ] \bigg)
    \end{aligned}
\end{equation}
with 
\begin{equation} \label{k_cat}
   K =  \operatorname{Tr}[\ \underbrace{a \ldots a}_m \ \sigma \ \underbrace{a^\dagger \ldots a^\dagger}_m ] \  = \quad   \operatorname{Tr}[\ \underbrace{a^\dagger \ldots a^\dagger}_{m} \ \underbrace{a \ldots a}_{m} \ \sigma  ]  
\end{equation}
By examination of Eqs. \ref{n_cat}-\ref{k_cat}, we notice in fact we only need to calculate 
\begin{equation}
  f(r) =  \operatorname{Tr}[\ \underbrace{a^\dagger \ldots a^\dagger}_{r} \ \underbrace{a \ldots a}_{r} \ \sigma  ]   
\end{equation}
for some generic $r$, and the \textit{signal-to-noise} ratio can be straightforwardly be derived from it through
\begin{equation}
    \Gamma_{c a t} = \frac{N_{c a t}-N_0}{\sqrt{N^2_{c a t}- (N_{c a t})^2}}=\frac{\frac{f(m+1)}{f(m)} - N_0}{\sqrt{\frac{f(m+1)+f(m+2)}{f(m)}-\bigg(\frac{f(m+1)}{f(m)}\bigg)^2}}
\end{equation}
where $$N_0 = \frac{1}{2}(\nu -1) $$
is the mean photon number of the initial thermal state (before applying the squeezing transformation). \\

Once the problem has been reduced to computing $f(r)$ using the method proposed in the previous section, we observe that the identification of perfect matchings of the ladder operators depends on the parity of $r$. We will examine both alternatives separately. \\

\textbf{Case $r$ even.} In this case, homogeneous matchings are possible (i.e, matchings that pair all of the creation operators among themselves, and all of the annihilation operators among themselves). 
\begin{itemize}
    \item The number of different homogeneous matchings that can be done is
    \begin{equation*}
        \bigg(\frac{r!}{2^{\frac{r}{2}}(\frac{r}{2})!}\bigg)^2
    \end{equation*}
    since there exist $r!$ different permutations of $r$ elements, but the $(\frac{r}{2})!$ permutations of the pairs are considered identical, and so are the $2^{\frac{r}{2}}$ permutations of the two elements within each pair. The square accounts for the fact that the same logic applies to both the creation and the annihilation operators. 
    \item If, on the contrary, we paired all of the creation operators with an annihilation one (all non-homogeneous pairs), we would have $r!$ possibilities (the first $a$ would choose among $r$ different $a^\dagger$'s, the second one among the remaining $r-1$, etc).
    \item The mixed approach in this case would be to make some $a^\dagger a$ pairs (necessarily an even number $s=2j$ of them) and mix the rest of the operators homogeneously. The number of options here takes the form:
    \begin{equation*}
       s! \bigg(\begin{array}{c}
   r   \\ s
\end{array} \bigg)^2 \ \bigg(
   \frac{(r-s)!}{2^{\frac{r-s}{2}}(\frac{r-s}{2})!}
 \bigg) ^2 \equiv b_r(s)
    \end{equation*}
\end{itemize}

Taking into account all these possibilities and applying Eqs.\ref{trace}, \ref{identities}, we obtain the expression:
\begin{equation}\label{reven}
\begin{aligned}
   f(r) & =  \sum_{j=0}^{\frac{r}{2}}   b_r(2j)   \cdot I_3^{2j} \cdot   I_1^{\frac{r-2j}{2}} \cdot  I_2^{\frac{r-2j}{2}}   \quad \quad \text{for $r$ even}
    \end{aligned}
\end{equation}

\textbf{Case $r$ odd.} In this case, homogeneous matchings are not possible, and an odd number of operators of each kind have to be paired non-homogeneously. Reasoning analogously as in the even case, we are left with:

\begin{equation}\label{rodd}
 \begin{aligned}
 f(r) & = \sum_{j=0}^{\frac{r-1}{2}}   b_r(2j+1) \cdot I_3^{2j+1} \cdot   I_1^{\frac{r-2j-1}{2}} \cdot   I_2^{\frac{r-2j-1}{2}} \quad \quad \text{for $r$ odd}
\end{aligned}
\end{equation}
Note that these expressions (Eqs.\ref{reven},\ref{rodd}) can be applied to compute the mean photon number of any Gaussian non-displaced state undergoing successive photon subtractions. However, unlike the case of kitten states (where the covariance matrix of the squeezed thermal state is diagonal), the equality $I_1 = I_2$ will not necessarily hold in general.
\section{Appendix E: Minimum Ergotropy of non-Gaussian states}\label{app:min_ergotropy_nongauss}
The non-Gaussian states that we propose in this work as most suitable batteries for precision energy harvesting have the following structure: 
\begin{equation} \label{ng_phadd_form}
\rho_{N G} = (a_1^\dagger)^m \ U_G \ \rho^0_{\text{th}} \ U_G^\dagger \ (a_1)^m \quad \text{for} \quad m = 1,2,3...
\end{equation}
where $m$ denotes their stellar rank, $U_G$ is a general Gaussian transformation, and $\rho^0_{\text{th}}$ is the passive thermal state of the system given some fixed fluctuation level parametrized by a temperature $T$. Figure 4 in the main text shows that, in principle, higher $m$ (that is, higher non-gaussianity) results in an advantage in terms of our figure of merit $\Gamma$. However, every photon added to the system conveys an energy increase, which may result in a violation of the constraint $\delta N \leq \epsilon$, and thus in the impracticability of the resulting state for the application that was pursued. In what follows, we will derive the explicit expression for such an increase. \\

The covariance matrix of $\rho^0_{\text{th}}$ is, for the single-mode case, given by 
$$ \sigma^0_{\text{th}} = \bigg(\begin{array}{cc}
   \nu  & 0 \\
   0  & \nu
\end{array} \bigg)\quad \quad \text{where} \quad \nu \equiv \coth \bigg(\frac{\omega}{2 T}\bigg)$$
Since $\rho^0_{\text{th}} $ is Gaussian, its mean photon number, which we denote $N_0$, can be directly calculated through \ref{expvalN_gaussian}: 
$$N_0 \equiv  N (\rho^0_{\text{th}}) =\frac{1}{2} (\nu -1) $$
And $\delta N =0$, since the state is passive (it is the initial input state). \\

Adding photons to this passive thermal state naturally increases the average energy. In particular, the least possible energy that one may reach after $m$ photon additions is that of the state $ \rho_{\text{th}}^{(+m)} = a^{\dagger^m}  \ \rho^0_{\text{th}}  \ a^m  $ (where we have dropped the indices since we are dealing with only one mode). That is, successively adding (analogous for subtracting) the $m$ photons to $\rho^0_{\text{th}}$, without previously performing any Gaussian transformations (which would increase energy even further). 
The analytical derivation of the mean photon number $\langle N \rangle = \langle a^\dagger a \rangle $ for this particular class of states $\rho_{\text{th}}^{(+m)}$ appears to be particularly simple. As before, this is equivalent to computing
\begin{equation}\label{trace_of_expvalN}
    \frac{1}{K}\operatorname{Tr}[\underbrace{a \ldots a}_{m \ \text{times}} \ \underbrace{a^\dagger a}_{N} \ \underbrace{a^\dagger \ldots a^\dagger}_{m \ \text{times}} \rho^0_{\text{th}}]
\end{equation}
with \begin{equation}
 K = \operatorname{Tr}[\underbrace{a \ldots a}_{m \ \text{times}} \ \underbrace{a^\dagger \ldots a^\dagger}_{m \ \text{times}} \rho^0_{\text{th}}]\end{equation}

For the identification and sum over all possible \textit{perfect matchings} of the ladder operators, we recall the diagonal form of the covariance matrix of $\rho^0_{\text{th}}$, and we realize there are only two non-zero entries: $$V_{11} = V_{22} = \nu$$
With this, $I_1 = I_2 = 0$, $I_3 = N_0 $, and $I_4 = N_0 +1 $, and we must only consider non-homogeneous perfect matchings $a^\dagger a$ or $a a^\dagger$. From a set of $m+1$ creations $a^\dagger$ and $m+1$ annihilators $a$ ordered as in Eq.\ref{trace_of_expvalN}, there are only two possible configurations: 
\begin{itemize}
    \item All pairs have the annihilator to the left: there are $m \cdot m!$ possible combinations for this, but $m!$ of them cancel out with the denominator $K$, so we are left with $$m \cdot \operatorname{Tr}[a a^\dagger] = m \cdot I_4 = m (N_0+1)$$
    \item The central pair $a^\dagger a$ is fixed, and the other operators can be paired as $aa^\dagger$ in $m!$ different ways, all $m!$ cancelled with the denominator $K$. In this case, we have just 
    $$ \operatorname{Tr}[a^\dagger a] =  I_3= N_0$$
\end{itemize}
Summing up both, we finally have the analytic expression for the mean photon number $N$ and photon number increase $\delta N$ of the state of lowest energy among those with the structure of Eq.\ref{ng_phadd_form}: 
\begin{equation}
     N (\rho_{\text{th}}^{(+m)}) = m(N_0+1) + N_0 
\end{equation}
\begin{equation}
    \delta N (\rho_{\text{th}}^{(+m)}) =m(N_0+1) = m \ \frac{\coth(\frac{1}{2T} ) + 1}{2}
\end{equation}

Therefore, the existence of a constraint $\epsilon$ in the maximum photon number input means we cannot apply as many photon additions as we want, but instead a maximum amount given by the expression: 
\begin{equation}
    m_{m a x}= \bigg\lfloor \frac{2 \epsilon}{\coth(\frac{1}{2T} ) + 1} \bigg\rfloor
\end{equation} \\

\begin{figure}[h!]
    \centering
    \includegraphics[width=0.6\linewidth]{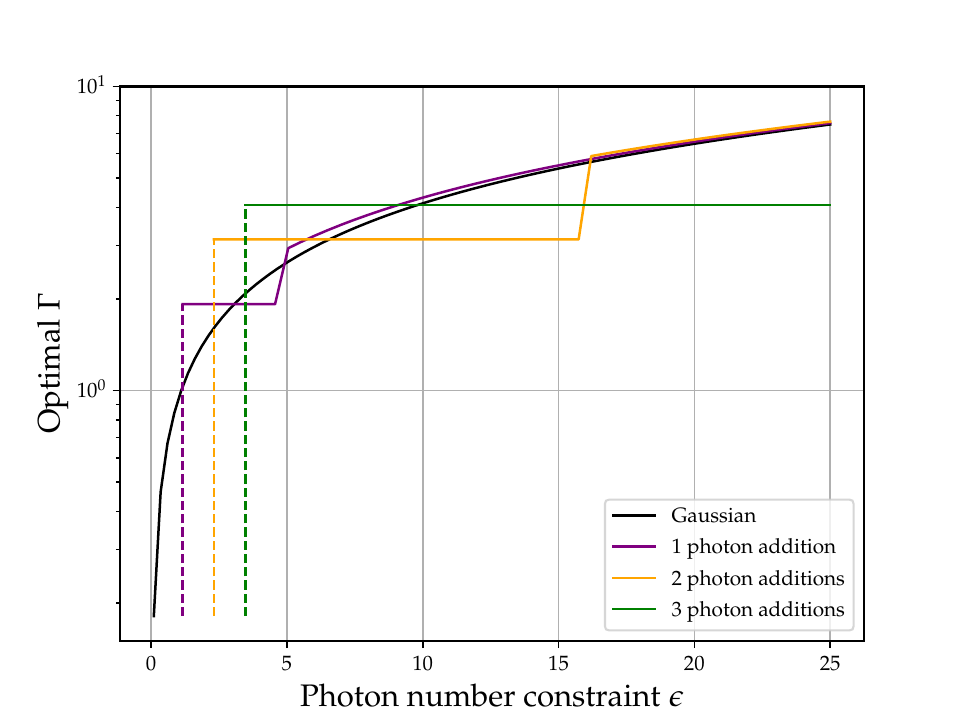}
    \caption{\raggedright\justifying Optimal $\Gamma$ for Gaussian and non-Gaussian photon-added single-mode states as a function of the constraint $\epsilon$, at fixed temperature $T =0.5 K$.} 
    \label{fig: optimal strategy}
\end{figure}

If some specific use-case has a very restrictive constraint $\epsilon$ on the maximum energetic input, the maximum allowed $m$ may be zero, and so the correct charging unitary to use in our application can only be a Gaussian one. In Fig.\ref{fig: optimal strategy}, one can see the optimal $\Gamma$ attainable with zero, one, two, and three photon-added states, as a function of the ergotropic constraint. The temperature has been set to $T = 0.5 K$ arbitrarily.
Non-Gaussian states are dominant from the initial region (where only Gaussian ones are feasible) onwards. However, one can identify an interval in which the Gaussian bound is widely violated by some photon-added state, while as the constraint is relaxed, the gain becomes close to negligible. This is consistent with the fact that, without the energetic restriction, one can achieve as high $\Gamma$ as desired just by applying displacement operations, which fall within the Gaussian realm.\\

\section{Appendix F: Effects of truncation}\label{app:truncation}
Despite the simplicity of the two particular cases outlined above, the exact calculations of expectation values on non-Gaussian states are, in general, computationally resource-demanding. They scale as $n!!$ with the number $n$ of ladder operators involved, counting both those included in the operator whose expectation value we are computing, and those applied to construct the non-Gaussian distribution. However, typical approaches to this exact calculation (i.e., discrete-variable approximations made by truncating the dimension of the Hilbert space) can lead to misleading and unfair conclusions, as we are about to show.  We will proceed by proving that truncating the Hilbert space of a coherent state $|\alpha \rangle$, allows the violation of the Gaussian bound. \\

Let us consider the Fock basis of a finite-dimensional Hilbert space with truncation to the second excited state: $\mathcal{H}= span(\{|0\rangle, |1\rangle, |2\rangle\})$. Then the definition of $|\alpha \rangle$ would be approximated to $|\alpha\rangle_t$ in the following way:
\begin{equation}
    |\alpha\rangle = \sum_n \frac{\alpha^{n}}{n!} e^{-\frac{|\alpha|^2}{2}} |n\rangle \approx  \underbrace{C \left( |0\rangle + \frac{\alpha}{\sqrt{1!}} |1\rangle + \frac{\alpha^2}{\sqrt{2!}} |2\rangle \right)}_{\equiv \ |\alpha\rangle_t} 
\end{equation}
where $C$ is a normalization factor required to make up for the lost probabilities:
\begin{equation}
    C = \left( 1 + |\alpha|^2 + \frac{|\alpha|^4}{2} \right)^{-1/2}
\end{equation}
Then, the photon number expectation value and uncertainty on $|\alpha\rangle_t$ yield:
\begin{equation}
   \Gamma(|\alpha\rangle_t)= \frac{\langle  N\rangle}{\Delta N}  = \frac{C^2 \left( |\alpha|^2 + |\alpha|^4 \right)}{\sqrt{C^2 \left( |\alpha|^2 + 2|\alpha|^4 \right) - C^4 \left( |\alpha|^2 + |\alpha|^4 \right)^2}}
\end{equation}
which can be shown to be always greater than $|\alpha|$, which is the existing bound on $\Gamma$ for coherent states, as shown in Eq.12 in the main text. 
In other words, by a mere approximation, we gain a non-classical resource in terms of charging precision. Obviously, by increasing the dimension of the Hilbert space enough, we should recover the correct result. However, this would make the calculations more tedious than the exact result we provide, employing the Wick theorem.

\section{Appendix G: Multimode case}\label{app:multimode}
\hyperref[lemma1]{Lemma~1} establishes the independency of the signal-to-noise ratio $\Gamma$ under Gaussian global (entangling) passive transformations in multimode Gaussian states. Below we provide a proof of this result:

\textit{Proof.} Consider a Gaussian state of $m$ modes $\rho$. Our figure-of-merit $\Gamma(\rho)$ is defined as the quotient between the difference in photon number in $\rho$ as compared to its thermal passive state $\rho_0$, and the photon number uncertainty $\Delta N (\rho)$: 
\begin{equation}
    \Gamma (\rho)= \frac{N(\rho) - N(\rho_0)}{\Delta N (\rho)}.
\end{equation}
Let $O$ be some generic passive-optics Gaussian transformation. Then $O \in O(2N) \cap Sp_{2n,\mathds{R}}$, i.e. $O$ is an orthogonal and symplectic matrix. By definition of passitivity, the application of $O$ cannot change the energy (nor the photon number) of state $\rho$, and thus the numerator is trivially not affected. As for the denominator: 
\begin{equation}
    \Delta N (\rho)= \sqrt{\langle N^2\rangle_\rho - N(\rho)^2 },
\end{equation}
where $N(\rho)^2$ is invariant since, as argued above, $N(\rho)$ is. The only term that could be possibly affected by the application of $O$ is then $\langle N^2\rangle_\rho$. Denoting the vector of quadratures $\bm{R} = (x_1,\ldots x_m, p_1, \ldots p_m)^T$, we expand $\langle N^2\rangle_\rho$ in terms of $\rho$'s covariance matrix $\sigma$ and displacement vector $\bm{\alpha}$:
\begin{equation}
    \langle N^2\rangle = \langle \bigg( \sum_{i=1}^m a^\dagger_i a_i \bigg)^2 \rangle = \langle \bigg( \sum_{i=1}^m \frac{1}{2}(x_i^2 + p_i^2 -1) \bigg)^2 \rangle =  \langle \bigg( \frac{1}{2}(\bm{R}^T \bm{R} -m) \bigg)^2 \rangle =   \frac{1}{4} ( \langle(\bm{R}^T \bm{R})^2\rangle -2m\langle \bm{R}^T \bm{R} \rangle +m^2) 
\end{equation}
By definition, $\bm{\alpha = \langle R\rangle}$, and $\sigma_{ij} = \frac{1}{2}(\{r_i,r_j\}),$ where $\bm{r} \equiv \bm{R}-\bm{\alpha}$, with $\langle \bm{r}\rangle =0$. Hence:
\begin{equation} \label{proof_eq1}
    \langle \bm{R}^T \bm{R} \rangle_\rho = \langle (\bm{\alpha} + \bm{r})^T (\bm{\alpha + \bm{r}) }\rangle = \langle \bm{\alpha}^T \bm{\alpha}\rangle + \langle \bm{\alpha}^T \rangle \langle\bm{r}\rangle + \langle \bm{r}^T \rangle \langle \bm{\alpha}\rangle + \langle \bm{r}^T \bm{r}\rangle \stackrel{\langle \bm{r}\rangle =0}{=} \langle \bm{\alpha}^T \bm{\alpha}\rangle  + \langle \bm{r}^T \bm{r}\rangle = ||\bm{\alpha}||^2  + \text{Tr}(\sigma)
\end{equation}
While the expectation value of its square yields:
\begin{equation}
    \langle (\bm{R}^T \bm{R})^2 \rangle = \langle (\bm{\alpha}^T \bm{\alpha})^2\rangle + 2\langle \bm{\alpha}^T \bm{r} \bm{r}^T \bm{\alpha}\rangle + \langle (\bm{\alpha}^T \bm{r})^2\rangle + \langle (\bm{r}^T \bm{\alpha})^2\rangle+ \langle (\bm{r}^T  \bm{r})^2\rangle +4 \langle (\bm{\alpha}^T \bm{\alpha})(\bm{\alpha}^T \bm{r})\rangle 
    +2 \langle (\bm{\alpha}^T \bm{\alpha})(\bm{r}^T \bm{r})\rangle 
    +4 \langle (\bm{\alpha}^T \bm{r})(\bm{r}^T \bm{r})\rangle 
\end{equation}
Terms in the previous expression with an odd number of $\bm{r}$'s cancel, and we are left with:
\begin{equation} \label{proof_eq2}
    \langle (\bm{R}^T \bm{R})^2 \rangle = ||\bm{\alpha}||^4 + 4 \bm{\alpha}^T \sigma \bm{\alpha} + \langle (\bm{r}^T  \bm{r})^2\rangle 
    +2 ||\bm{\alpha}||^2 \text{Tr}(\sigma)
\end{equation}
Employing Wick's theorem, the term $\langle (\bm{r}^T  \bm{r})^2\rangle$ can be expressed as
\begin{equation} \label{proof_eq3}
    \langle (\bm{r}^T  \bm{r})^2\rangle = 2 \text{Tr}(\sigma^2) + (\text{Tr} (\sigma))^2 -m 
\end{equation}

By inspection of Eqs.\ref{proof_eq1}-\ref{proof_eq3}, it suffices to prove that $||\bm{\alpha}||$, $\bm{\alpha}^T \sigma \bm{\alpha}$, $\text{Tr}(\sigma)$, and $\text{Tr}(\sigma^2)$ are invariant under the application of passive optics transformations:
\begin{equation}
\left.
\begin{aligned}
\bm{\alpha} \rightarrow O \bm{\alpha}  \\
\sigma \rightarrow O \sigma O^T \\
\text{$O$ is orthogonal}, OO^T = \Id
\end{aligned}
\right\}
\;\Longrightarrow\;
\begin{cases}
||O \bm{\alpha}|| = ||\bm{\alpha} ||, \\
(O \bm{\alpha})^T (O\sigma O^T) (O \bm{\alpha}) =  \bm{\alpha}^T O^T (O\sigma O^T) O \bm{\alpha}  = \bm{\alpha}^T \sigma \bm{\alpha},\\
\text{Tr}(O \sigma O^T) = \text{Tr}(O^T O \sigma) = \text{Tr}(\sigma) ,\\
\text{Tr}((O \sigma O^T)^2) =
\text{Tr}((O \sigma O^T)(O \sigma O^T)) = \text{Tr}(O \sigma^2 O^T) = \text{Tr}(\sigma^2)~.
\end{cases}
\end{equation}
Since all of its components are invariant under $O$, $\Gamma(\rho)$ also remains unchanged.
\hfill \qed \\ \\

\begin{figure}[h!]
    \centering
    \includegraphics[width=0.6\linewidth] {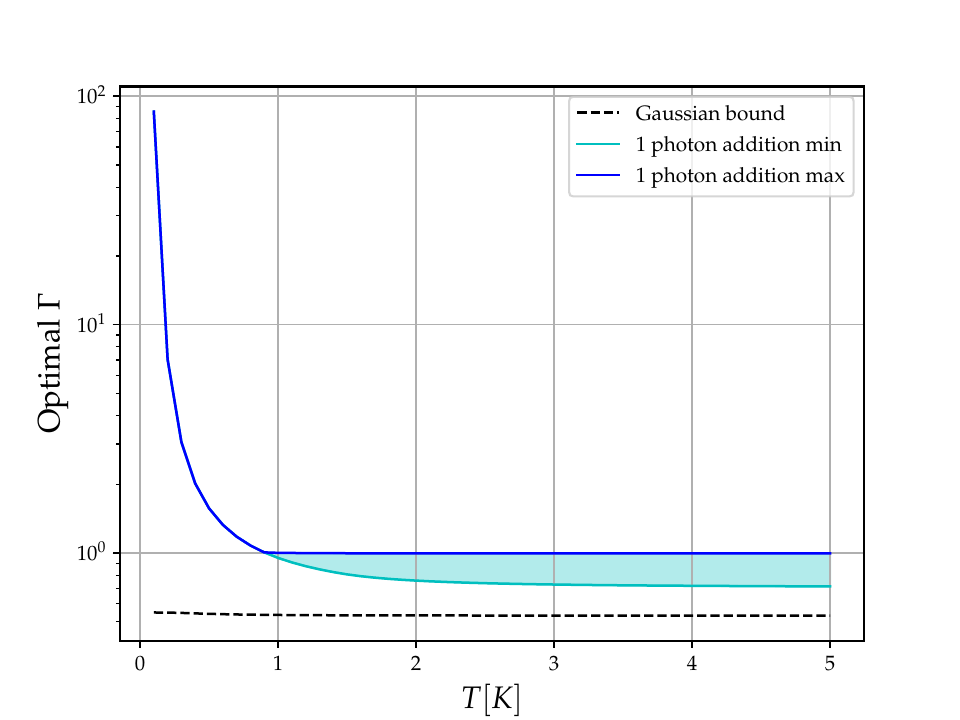}
    \caption{\raggedright\justifying Optimal \textit{signal-to-noise} ratio for Gaussian, one-photon added separable and one-photon added entangled states of two modes, as a function of system's temperature. The energetic constraint has been set as $\sum_{i=1,2} \delta N_i \leq \epsilon = 5 $}
    \label{entanglement_twomodes}
\end{figure}

In Fig.\ref{entanglement_twomodes} we show the value of $\Gamma$ for the optimal two-mode Gaussian state (dashed black), and one-photon added state, both in the separable case (light blue, yielding the worst ratio) and the maximally entangled one (dark blue). The shaded area represents the gain given by the application of a $\theta = \pi/4$ beam splitter. Note that, at low temperatures, entanglement plays no role, since, as mentioned above, in the pure-state regime, Fock states are always optimal in terms of photon-number uncertainty. 
 
\begin{figure}[h]
    \centering
    \includegraphics[width= 0.6\linewidth] {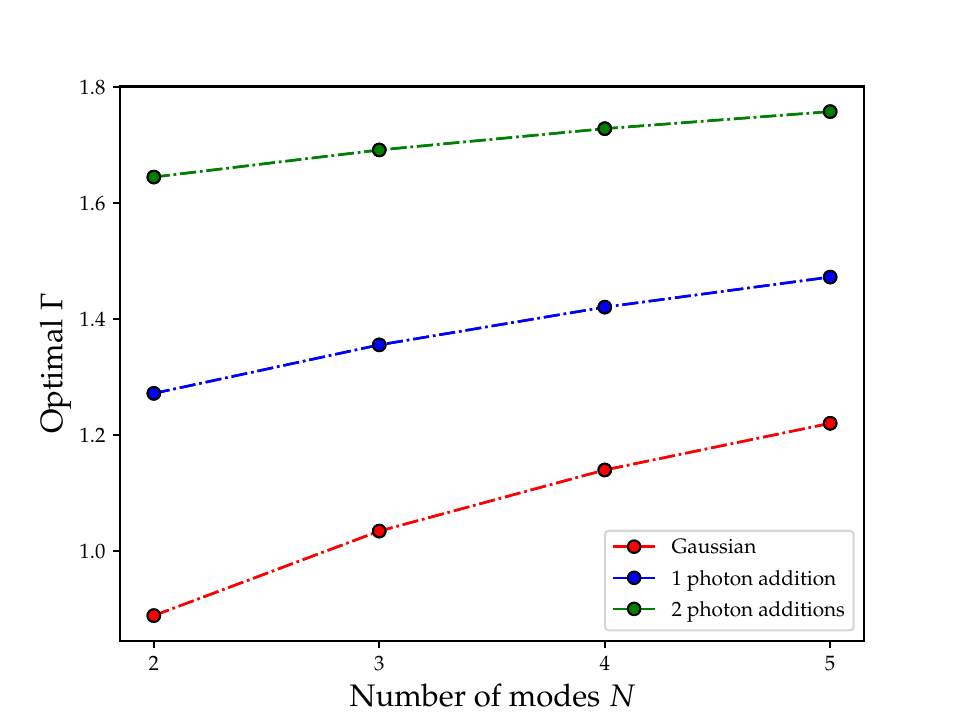}
    \caption{\raggedright\justifying Optimal \textit{signal-to-noise} ratio for Gaussian, one-photon added and two-photon added entangled states as a function of the number of modes $N \in \{2,3,4,5\}$. System's temperature and energetic constraint have been set at $T=0.5K$ and $\sum_i \delta N_i \leq \epsilon = 10 $, respectively.}
    \label{fig: many_modes}
\end{figure}

Finally, we address the question of whether this advantage is extensive with the number of modes. In Fig.\ref{fig: many_modes} we show the optimal \textit{signal-to-noise} ratio $\Gamma$ (upon tuning of the squeezing, beam-splitting, phase-shifting and displacement parameters) attainable for multimode batteries of increasing degrees of non-Gaussianity. From the results, we conclude that the hierarchy
in extractable \textit{signal-to-noise} ratios established by the degree of non-Gaussianity in the single mode case
is a property that can be generalized to larger $N$, and the existent increasing trend of $\Gamma$ with $N$ suggests that bigger and more complex non-gaussian battery systems
hold a promising potential as highly precise energy storage devices.

\section{Appendix H}\label{app:noise_detection}
\subsection{1. SNR under lossy/noisy detection}\label{app:monotoneSNR}
Let $\rho$ be the photonic output state with $\delta N=\langle N\rangle-N_0$ and $(\Delta N)^2=\mathrm{Var}_\rho(N)$, and define the \emph{state-level} SNR $\Gamma=\delta N/\Delta N$. We model the detection/readout as a linear loss channel of efficiency $\eta\in[0,1]$ (binomial thinning of photon counts) followed by additive, classical readout noise $B$ that is independent of $N$, with mean $\beta$ and variance $\sigma_B^2$. Applying the same map to both the initial and the charged state gives
\begin{align}
\langle N_{\rm det}\rangle &= \eta\,\langle N\rangle,\\
\mathrm{Var}(N_{\rm det}) &= \eta^2\,\mathrm{Var}(N)\;+\;\eta(1-\eta)\,\langle N\rangle\;+\;\sigma_B^2,
\end{align}
so that the \emph{measured} SNR is
\begin{equation}
\Gamma_{\rm det}
=\frac{\delta N_{\rm det}}{\sqrt{\mathrm{Var}(N_{\rm det})}}
=\frac{\eta\,\delta N}{\sqrt{\eta^2\Delta N^2+\eta(1-\eta)\langle N\rangle+\sigma_B^2}}.
\end{equation}
Since $\eta(1-\eta)\langle N\rangle+\sigma_B^2\ge 0$, one has $\mathrm{Var}(N_{\rm det})\ge \eta^2\Delta N^2$, which implies the \emph{monotonicity} (data-processing) inequality
\begin{equation}
\Gamma_{\rm det}\ \le\ \Gamma,
\end{equation}
with equality if and only if $\eta=1$ and $\sigma_B^2=0$ (ideal unit-efficiency, noiseless readout). As an illustration, for the pure Fock state $\ket{1}$ one has $\Delta N=0$ and hence state-level $\Gamma=\infty$, whereas after loss/noise the detected variable is Bernoulli (plus $B$) with variance $\eta(1-\eta)+\sigma_B^2$, yielding a finite $\Gamma_{\rm det}$. This clarifies why experimental tests must compare $\Gamma_{\rm det}$ to \emph{detection-propagated} bounds (classical or Gaussian) obtained by pushing the corresponding state-level bounds through the same loss/noise map.

For $T=1\,\mathrm{K}$ ($\nu\simeq1.313$, $N_0\simeq0.157$), energy budget $\epsilon=5$, overall efficiency $\eta=0.60$, and electronic noise $\sigma_B=0.30$, the propagated Gaussian \emph{bound} gives $\Gamma_{G,\mathrm{det}}\approx1.84$. An optimised single photon–added state (with $\Gamma\simeq1.6\,\Gamma_G\approx4.16$) attains $\Gamma_{\mathrm{det}}\approx2.21>\Gamma_{G,\mathrm{det}}$, thus violating the bound at the electrical readout.

\subsection{2. Finite-time TUR for the electrical readout}\label{app:TUR_SM}
We evaluate precision on the integrated charge $Q_\tau=\int_0^\tau I(t)\,dt$ delivered to a resistive load at (effective) temperature $T_e$ over a window $\tau$ (accumulate-and-sample). The measured SNR is
\begin{equation}
\Gamma_\tau\ :=\ \frac{\langle Q_\tau\rangle-\langle Q_\tau\rangle_0}{\sqrt{\mathrm{Var}(Q_\tau)}}\,.
\end{equation}
In stochastic thermodynamics, the finite-time thermodynamic uncertainty relation (TUR) for an integrated current yields
\begin{equation}\label{eq:TUR_main}
\frac{\mathrm{Var}(Q_\tau)}{\langle Q_\tau\rangle^2}\ \ge\ \frac{2}{\Sigma_\tau}\,,
\qquad
\Sigma_\tau=\frac{1}{k_B T_e}\int_0^\tau \langle P_{\rm diss}(t)\rangle\,dt,\quad
P_{\rm diss}(t)=I(t)^2R,
\end{equation}
where $\Sigma_\tau$ is the total (dimensionless) entropy production in the load during $\tau$ \cite{BaratoSeifert2015_TUR,Gingrich2016_DissipationBounds,DechantSasa2018_EntropicBound,KoyukSeifert2019_TimeDependentTUR,HorowitzGingrich2020_NatPhys_TURReview}. For stationary operation with mean current $\langle I\rangle$, one has $\Sigma_\tau=(\langle I\rangle^2R/k_BT_e)\,\tau$, so
\begin{equation}\label{eq:TUR_ceiling}
\Gamma_\tau\ \le\ \sqrt{\Sigma_\tau/2}\ =\ \sqrt{\frac{\langle I\rangle^2 R\,\tau}{2k_B T_e}}\,.
\end{equation}
This ceiling is \emph{detector-level} and independent of how the optical state is prepared. Optical non-classicality/non-Gaussianity can reduce $\mathrm{Var}(Q_\tau)$ at fixed mean, thereby pushing $\Gamma_\tau$ closer to the same ceiling; our witnesses (Theorems~\ref{thm:A}–\ref{thm:B}) remain state-level and are tested by comparing $\Gamma_{\rm det}$ to detection-propagated bounds (See \hyperref[app:monotoneSNR]{Appendix H.1} above). The TUR statement only supplies an absolute, dissipation-limited benchmark at the harvester.

\end{document}